\documentclass[cup6b]{cupbook}
\usepackage{color,graphicx,subfigure,graphics,epsfig,amsmath,amssymb}
\usepackage{symb}
\usepackage[round, comma]{natbib}
\usepackage{mathrsfs}
\usepackage{url}
\def\deg2{{~\mathrm{deg}^2}}
\def\A{{\AA~}}
\def\om{\Omega_m}

\def\({\left(}
\def\){\right)}
\def\intzinf{\int_z^\infty}
\def\be{\begin{equation}}
\def\ee{\end{equation}}
\def\bea{\begin{eqnarray}}
\def\eea{\end{eqnarray}}

\def\name{{\em Fisher4Cast}}

\def\om{\Omega_m}

\def\({\left(}
\def\){\right)}

\def\bao{Baryon Acoustic Oscillations}
\def\intzinf{\int_z^\infty}

\begin{document}
\def\newblock{\hskip .11em plus .33em minus .07em}
\maketitle
\author[Bassett and Hlozek, 2009]{Bruce A. Bassett $^{1,2,a}$ \& Ren\'{e}e Hlozek$^{1,2,3,b}$
\and
\\ $^1$ South African Astronomical Observatory,\\  ~~~~ Observatory, Cape Town, South Africa 7700 \\
\\ $^2$ Department of Mathematics and Applied Mathematics, \\ ~~~~University of Cape Town, Rondebosch, Cape Town, South Africa 7700 \\
\\ $^3$ Department of Astrophysics, University of Oxford \\ ~~~~Keble Road, Oxford, OX1 3RH, UK
\\$^a$ bruce@saao.ac.za $~~^b$ renee.hlozek@astro.ox.ac.uk }
\chapter[Baryon Acoustic Oscillations]{{\bf Baryon Acoustic Oscillations} \nonumber}
\addvspace{2.5in }
\begin{abstract}
 \bao~(BAO) are frozen relics left over from the pre-decoupling universe. They are the standard rulers of choice for 21st century cosmology, providing distance estimates that are, for the first time, firmly rooted in well-understood, linear physics. This review synthesises current understanding regarding all aspects of BAO cosmology, from the theoretical and statistical to the observational, and includes a map of the future landscape of BAO surveys, both spectroscopic and photometric\footnote{This review is an extended version of a chapter in the book {\em Dark Energy} Ed.~Pilar Ruiz-Lapuente, Cambridge University Press (2010, ISBN-13: 9780521518888)}.
\end{abstract}
\section{Introduction}
Whilst often phrased in terms of the quest to uncover the nature of dark energy, a more general rubric for cosmology in the early part of the 21st century might also be the ``the distance revolution". With new knowledge of the extra-galactic distance ladder we are, for the first time, beginning to accurately probe the cosmic expansion history beyond the local universe. While standard candles -- most notably Type Ia supernovae (SNIa) -- kicked off the revolution, it is clear that Statistical Standard Rulers, and the \bao~(BAO) in particular, will play an  increasingly important role.

In this review we cover the theoretical, observational and statistical aspects of the BAO as standard rulers and examine the impact BAO will have on our understanding of dark energy, and the distance and expansion ladder. Fisher matrix forecasts for BAO surveys can be easily computed using the publically released, GUI-based, \name~code, which is publically available\footnote{http://www.cosmology.org.za}.
\subsection{A Brief History of Standard Rulers and the BAO}
Let us start by putting the BAO in context. The idea of a standard ruler is one familiar from everyday life. We judge the distance of an object of known length (such as a person) by its angular size. The further away it is, the smaller it appears. The same idea applies in cosmology, with one major complication: space can be curved. This is similar to trying to judge the distance of our known object through a smooth lens of unknown curvature. Now when it appears small, we are no longer sure it is because it is far away. It may be near and simply appear small because the lens is distorting the image. This degeneracy between the curvature of space and radial distance has not been the major practical complication in cosmology over the past century, however. That honour goes to a fact that has plagued us since the beginning of cosmology: we don't know how big extragalactic objects are in general, in the same way that we don't know how bright they intrinsically are. This problem was at the heart of the great debate between Shapley and Curtis over the nature of galaxies. Shapley argued that they were small and inside our own galaxy while Curtis maintained that they were extragalactic and hence much larger.

To be useful for cosmology, we need a {\em standard ruler}: an object of a known size at a single redshift, $z$, or a population of objects at different redshifts whose size changes in a well-known way (or is actually constant) with redshift. Ideally the standard ruler falls into both classes, which, as we will argue below, is the case for the BAO, to good approximation. BAO are however a new addition to the family of putative standard rulers. A few that have been considered in the past include ultra-compact radio sources \cite{kellermann93,gurvits94} which indeed lead in 1996, prior to the SNIa results, to claims that the density of dark matter was low, $\Omega_m < 0.3,$ with a non-zero cosmological constant of indeterminate sign \cite{jackson97}. Another radio standard ruler candidate is provided by double-lobed radio sources \cite{buchalter_radio}. These Fanaroff-Riley Type IIb radio galaxies were suggested as cosmological probes as early as 1994 \cite{daly_friib}, and subsequent analyses have given results consistent with those from SNIa \cite{daly, latest_daly}. An alternative approach uses galaxy clusters. Allen {\em et al.} relate the X-ray flux to the cluster gas mass, and in turn, its size, providing another standard ruler under the assumption that the gas fraction is constant in time. This too leads to results consistent with those from SNIa \cite{allen_schmidt, latest_allen}.

Beyond this we move into the realm of Statistical Standard Rulers
(SSR), of which BAO are the archetype. SSR exploit the idea that the
clustering of galaxies may have a preferred scale in it which, when
observed at different redshifts, can be used to constrain the angular
diameter distance. The idea of using a preferred clustering scale
to learn about the expansion history of the cosmos has a fairly long
history in cosmology, dating back at least to 1987 and perhaps
earlier. In their conclusions Shanks {\em et al.} \cite{shanks87} forsee:
\begin{quote}
{\em There is one further important reason for searching for weak
features in $\xi_{qq}(r),$ the quasar-quasar correlation function, at
large separations. If a particular feature were found to appear in
both the galaxy correlation function at low redshift and the QSO
correlation function at high redshift, then a promising new
cosmological test for $q_0$ might be possible.}
\end{quote}
A series of later analyses further built up the idea of SSR in
cosmology using variously as motivation the turn-over in the power
spectrum due to the transition from radiation to matter-domination,
the mysterious $128 h^{-1} \mathrm{Mpc}$ feature detected in early pencil-beam
surveys \cite{broadhurst_ellis} and the realisation that inflation could inject
a preferred scale into the primordial power spectrum. These early
studies often found tentative evidence for a low-density universe
and/or non-zero cosmological constant e.g. \cite{deng,broad_jaffe_ssr}. Since the preferred scale could not be
accurately predicted {\em a priori} these studies only provided the
relative size of the SSR at different redshifts. Nevertheless, it was realised that this could provide interesting constraints on the expansion history of the universe \cite{roukema9911413,roukema2_0106135}.

\begin{figure}[htbp!]
\begin{center}
\includegraphics[width = 3.8in, height = 3.1in]{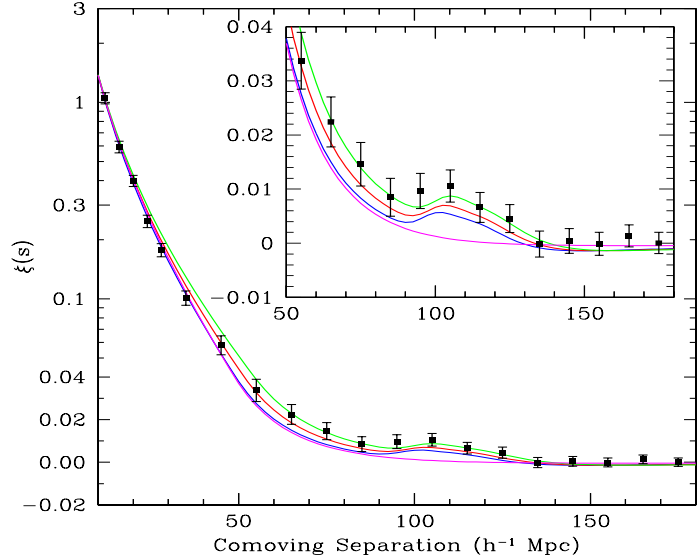}
\caption{The Baryon Acoustic Peak (BAP) in the correlation function -- the BAP is visible in the clustering of the SDSS LRG galaxy sample, and is sensitive to the matter density (shown are models with $\Omega_mh^2=0.12$ {\bf (top)}, 0.13 {\bf (second)} and 0.14 {\bf (third)}, all with $\Omega_bh^2=0.024$). The bottom line without a BAP is the correlation function in the pure CDM model, with $\Omega_b=0$. From Eisenstein {\em et al.,}  2005 \cite{eisenstein_05}.\label{eisenstein:fig}}
\end{center}
\end{figure}

\begin{figure}[htbp!]
\begin{center}
\includegraphics[width = 3.8in, height = 3.1in]{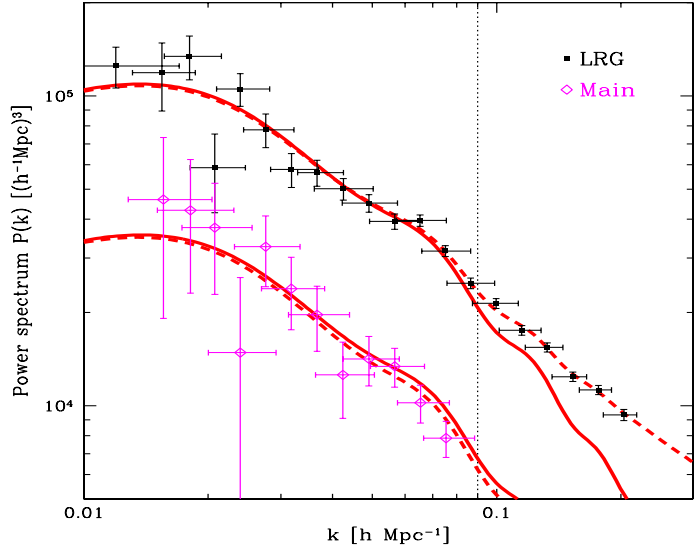}
\caption{Baryon Acoustic Oscillations (BAO) in the SDSS power spectra -- the BAP of the previous figure now becomes a series of oscillations in the matter power spectrum of the SDSS sample. The power spectrum is computed for both the main SDSS sample {\bf (bottom curve)} and the LRG sample {\bf (top curve)}, illustrating how LRGs are significantly more biased than average galaxies. The solid lines show the $\Lambda$CDM fits to the WMAP3 data \cite{wmap3}, while the dashed lines include nonlinear corrections. Figure from Tegmark {\em et al.,} 2006 \cite{tegmark_lrg}. \label{sdss:fig}}
\end{center}
\end{figure}

BAO entered the fray initially as a putative explanation for the apparent excess clustering
around $100 h^{-1} \mathrm{Mpc}$  but were found to be too weak to be the
origin for the apparent excess \cite{eisenstein_hu_1998, meiskin}.
The idea of using BAO themselves to learn about cosmological parameters seems to date first from Eisenstein {\em et al.} (1998) \cite{eis98_complimentarity} who wrote:
\begin{quote}
{\em Detection of acoustic oscillations in the matter power spectrum
would be a triumph for cosmology, as it would confirm the standard
thermal history and the gravitational instability paradigm. Moreover,
because the matter power spectrum displays these oscillations in a
different manner than does the CMB, we would gain new leverage on
cosmological parameters.}
\end{quote}
The first photometric proposal for using the BAO as standard rulers for learning
about cosmology appears to date from 2001 \cite{cooray_halo}. The real foundations, however, of the modern ideas on BAO, their detection and use, were laid by Eisenstein (2003) \cite{eisenstein2003}, Blake and Glazebrook  (2003) \cite{blake_glaze_simulation} and a slew
of later papers \cite{hu_haiman,amendola_de, blake_bao, glaze_blake, white_bao, wang_bao, huff, angulo, komatsu_angular_bao, parkinson06, dunemartin} which developed hand-in-hand with the analysis of real
data.  Tantalising hints for the existence of the BAO were already
visible in the Abell cluster catalogue \cite{miller}, but
definitive detections had to wait for the increased survey volume and
number density of galaxies achieved in the SDSS and 2dF redshift surveys, which
immediately yielded strong constraints on both curvature and dark energy at $z < 0.5$
\cite{eisenstein_05,cole_2df,tegmark_lrg, percival_lrg, hui_bao}. Figures
(\ref{eisenstein:fig}) and (\ref{sdss:fig}) show the original evidence
for the acoustic signature in the correlation function and power
spectrum.  Extracting the BAO scale from the matter power spectrum
remains a thriving area of research in contemporary cosmology, as we
discuss later in Section \ref{experiments} on current and future BAO
surveys.

\subsection{Cosmological Observables}
We now discuss the relevant cosmological observables that are derived from standard rulers in general, and the BAO in particular. The \bao~in the radial and tangential directions provide measurements of the Hubble parameter and angular diameter distance respectively. The Hubble parameter, $H \equiv \dot{a}/a$ -- where $a$ is the scale factor of the universe -- can be written in dimensionless form using the Friedmann equation as \be E(z) \equiv \frac{H(z)}{H_0} = \sqrt{ \Omega_m(1+z)^3 + \Omega_{\mathrm{DE}}f(z) + \Omega_k(1+z)^2 + \Omega_{rad}(1+z)^4}\,, \label{hubble_eq}
\ee where $f(z)$ is the dimensionless dark energy density and $\Omega_k = -\frac{k}{H_0^2a^2} = 1- (\Omega_m + \Omega_{DE} + \Omega_{rad})$ is the density parameter of curvature with $\Omega_k = 0$ corresponding to a flat cosmos. $\Omega_{m},\Omega_{rad}$ are the matter and radiation densities with corresponding equations of state $w_i \equiv p_i/\rho_i = 0,\frac{1}{3}$ for $i=m,rad$ respectively.

If one treats the dark energy as a barotropic fluid with an equation of state with arbitrary redshift dependence, $w(z)$, the continuity equation can be directly integrated to give the evolution of the dimensionless dark energy density, $f(z)= \rho_{DE}/\rho_{DE}(z=0)$, via \be f(z) = \exp\left[3 \int_0^z  \frac{1+w(z')}{1+z'}dz'\right]\,. \label{fz} \ee
When we quote constraints on dark energy it will typically be in terms of the CPL parameterisation \cite{cp,linder_w}
\be w(z) = w_0 + w_a\frac{z}{1+z}\,, \label{wcpl} \ee which has \be f(z) = (1+z)^{3(1+w_0+w_a)} \exp \left\{-3w_a \frac{z}{1+z}\right\}\,.\label{feq} \ee

Much of the quest of modern cosmology is to constrain the allowed range of $w(z)$ (or $f(z)$) and hence use this to learn about physics beyond the standard model of particle physics and General Relativity. Apart from direct measurements of the Hubble rate, one of the ways to constrain $w(z)$ using cosmology is through distance measurements. Core to defining distances in the FLRW universe is the dimensionless, radial, comoving distance: \be \chi(z) \equiv\int_0^z \frac{dz'}{E(z')}\,. \label{chiz}\ee
One then builds the standard cosmological distances using $\chi(z)$ to give the {angular diameter distance}, $d_A(z)$ via
\be d_A(z) = \frac{c}{H_0 (1+z) \sqrt{-\Omega_k}}\sin\left(\sqrt{-\Omega_k}\chi(z)\right) \label{da_eq}
\ee and the {luminosity distance}, $d_L(z)$, given via the distance duality as \be  d_L(z) = (1+z)^2 d_A(z)  \label{dist_duality} \ee

\begin{figure}[htbp!]
\begin{center}
\includegraphics[width = 2in, height = 1.6in]{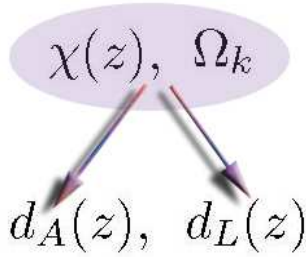}
\caption{Curvature and $\chi(z)$ define cosmological distances -- In a flat Universe, the cosmological distances are determined by $\chi(z) \propto \int_0^z dz'/E(z').$ In a general FLRW model, however, spatial curvature bends the light rays away from straight lines and hence alters distances, meaning that one needs to know both $\Omega_k$ and $\chi(z)$. As a result distance measurements always show a degeneracy between curvature ($\Omega_k$) and dynamics ($H(z)$). \label{schematic_dist}}
\end{center}
\end{figure}
The expression (Eq.~(\ref{da_eq})) for $d_A(z)$ holds for all values of the curvature, $\Omega_k$, since for $\Omega_k < 0$ the complex argument in Eq.~(\ref{da_eq}) converts the $\sin$ function to the $\sinh$ function. Hence the two key quantities that determine distances in cosmology are the dimensionless distance $\chi(z)$ and $\Omega_k$, shown schematically in
Figure~(\ref{schematic_dist}). The link (Eq.~(\ref{dist_duality})) between $d_A(z)$ and $d_L(z)$ holds in any metric theory of gravity as long as photon number is conserved. This distance duality can be tested and used to look for exotic physics \cite{bassett_distance2, bassett_distance, uzan_dd, more_dd, verde_dd}.

\begin{figure}[htbp!]
\begin{center}
\includegraphics[width = 3.8in, height = 3.1in]{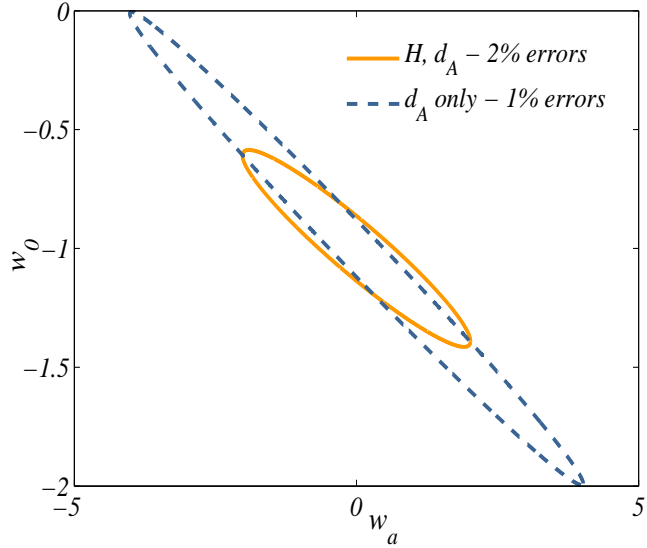}
\caption{Breaking the curvature-dark energy degeneracy -- Error ellipses for the CPL parameters $w_0, w_a$, in two survey scenarios after marginalising over curvature: one consisting of two measurements at $z = 1$ and $z=3$ of the angular diameter distance, with $1\%$ errors on $d_A(z)$ (blue dashed curve), and another of measurements of $H(z)$ and $d_A(z)$ (solid orange curve) where the errors have been doubled for both observables. In both surveys we assume a prior on curvature of 30, and $\mathrm{Prior}(\Omega_m) = \mathrm{Prior}(H_0) = 1000.$ Even given a weak prior on curvature, combining measurements from multiple probes helps break the curvature-dark energy degeneracy. Figure produced using \name.\label{hda_curv}}
\end{center}
\end{figure}

Distances have a significant disadvantage over pure Hubble measurements: they require an integral over $f(z)$ which is itself an integral over $w(z)$. Hence, any interesting features in $w(z)$ tend to be washed out in distance measurements. There is also another problem: if we look at Eq.~(\ref{da_eq}) for $d_A(z)$ we notice that if we make no assumptions about $f(z)$, then even perfect distance measurements cannot break the degeneracy between $f(z)$ and $\Omega_k$ \cite{weinberg1972}. This is not a fundamental problem if one assumes that $w(z)$ has finite degrees of freedom, e.g. in Eq.~(\ref{wcpl}), but one must remember that the degeneracy is being broken artificially by hand through ones choice of parameterisation and not by the data. As an example, consider Figure~(\ref{hda_curv}), which shows two possible dark energy survey configurations; one with $1\%$ measurements of the angular diameter distance at redshifts of $z=1,3,$ and another with measurements of both the Hubble parameter and angular diameter distance, but with double the errors on each observable. Even given the increase in the error on the observables, the dark energy constraints are significantly improved when including data from these complimentary probes when marginalising over curvature.
In principle this degeneracy can be broken even with arbitrary $f(z)$ by simultaneous measurements of both Hubble and distance; one can explicitly write $\Omega_k$ in any FLRW model (with no recourse to the Einstein field equations) as \cite{CCB,de_degen}:
\be
\Omega_k = \frac{[H(z)D'(z)]^2-H_0^2}{[H_0D(z)]^2}\,\label{curv_ok},
\ee
where $D$ is the dimensionless, transverse, comoving distance with $D(z) = (c/H_0) (1+z) d_A(z)\,.$
This relation gives the value of $\Omega_k$ {\em today}, as a function of measurements at {\em any} redshift $z$. Hence this can be turned into a powerful test of the Copernican Principle \cite{copernican}. Since $\Omega_k$ is a single number, the right hand side will have the same value when measured at any redshift if one is in a FLRW background. If it is found to vary with redshift, then we do not live in a FLRW universe.

The beauty of \bao~is that they provide both $d_A(z)$ and $H(z)$ using almost completely linear physics (unlike SNIa for example which involve highly complex, nonlinear, poorly understood stellar explosions). In addition, they offer the as yet unproven possibility of delivering constraints on growth though the change in the amplitude of the power spectrum. The time-dependence of the matter density perturbations, $\delta \rho/\rho$ obeys the equation
\begin{equation}
\ddot{\delta} + 2H\dot{\delta} = 4\pi G \rho_m \delta.
\label{deltaeq}
\end{equation}
The time-dependence of the growing mode of this equation is given by the growth function, $G(z)$ which in a flat, $\Lambda$CDM model satisfies \cite{growth_eis}:
\begin{equation}
G(z) = \frac{5 \om E(z)}{2} \intzinf \frac{(1+z')dz'}{E(z')^3}\,,
\label{growthz}
\end{equation} while in a general universe with curvature and dark energy dynamics Eq.~(\ref{deltaeq}) can be rewritten as \cite{fisher_release}:
\be
\delta'' + \frac{3}{2}\left(1 +\frac{\Omega_k(x)}{3} - w(x)\Omega_{\mathrm{DE}}(x)\right)\frac{\delta'}{x} -\frac{3}{2}\Omega_m(x)\frac{\delta}{x^2} = 0, \label{newg}
\ee
in terms of the dimensionless scale factor $x = a/a_0 = 1/(1+z),$ where $a_0 = c/H_0(\sqrt{\Omega_k})$ is the radius of curvature of the universe. From Eq.~(\ref{growthz}) we can see that the growth contains a mixture of Hubble and ``distance" information - as a result measurements of growth are potentially powerful probes of dark energy.

\subsection{Statistical Standard Rulers\label{stat}}
To illustrate the idea underlying Statistical Standard Rulers (SSR), imagine that all galaxies were positioned at the intersections of a regular three-dimensional grid of known spacing $L$. Measuring angular diameter distances as a function of redshift would be trivial in this case (c.f. Eq.~(\ref{da_eq})) and we would also have the expansion rate as a function of redshift, measured at a discrete set of redshifts corresponding to the mid-points between galaxies. Now imagine that we start to randomly insert galaxies into this regular grid. As the number of randomly distributed galaxies increases the regular grid pattern will rapidly become hard to see by eye. However, the underlying grid pattern would still be detectable statistically, for example in the Fourier transform.

\begin{figure*}[htbp!]
\begin{center}
$\begin{array}{@{\hspace{-0.6in}}c@{\hspace{-0.3in}}c}
\includegraphics[width = 3.5in,height = 3in]{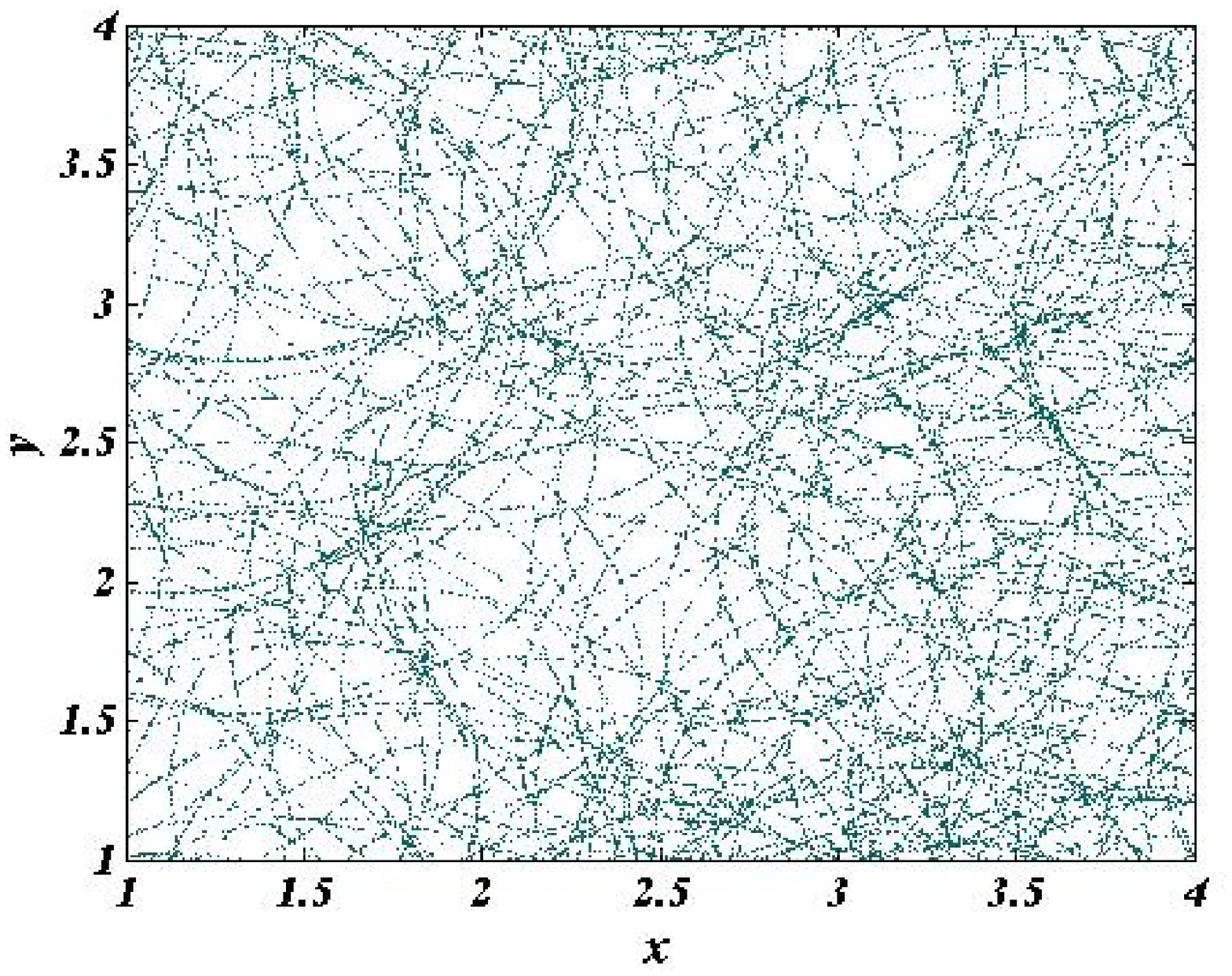} &
\includegraphics[width = 3.5in, height = 3in]{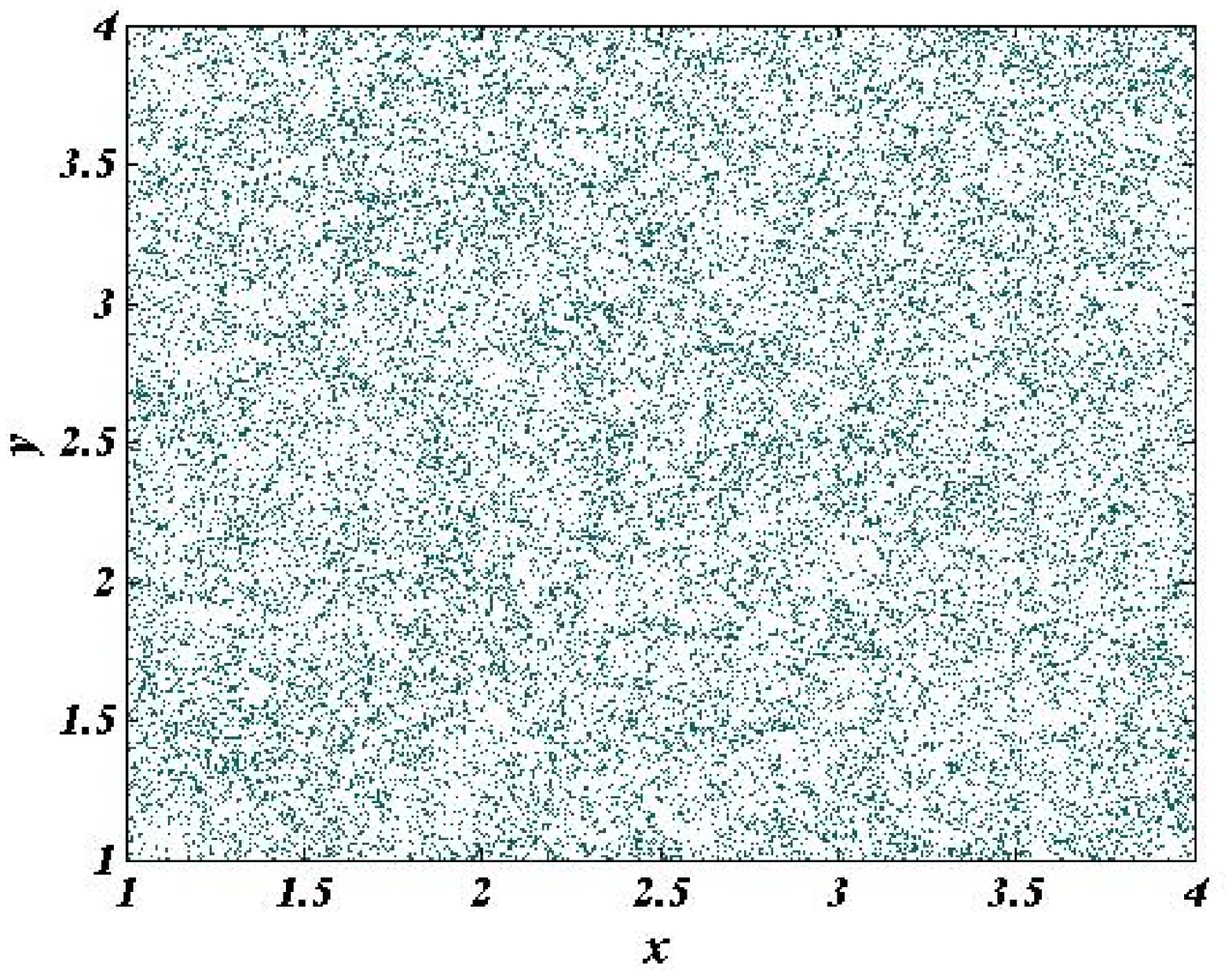} \\ [0.0cm]
 \end{array}$
  \caption{Rings of power superposed. Schematic galaxy distribution formed by placing the galaxies on rings of the same characteristic radius $L.$ The preferred radial scale is clearly visible in the left hand panel with many galaxies per ring. The right hand panel shows a more realistic scenario - with many rings and relatively few galaxies per ring, implying that the preferred scale can only be recovered statistically. \label{rings:fig}}
  \end{center}
 \end{figure*}

However, a regular grid distributed throughout space would provide an absolute reference frame and would break the continuous homogeneity of space down to a discrete subgroup (formed by those translations which are multiples of the grid spacing $L$). To get to the core of SSR consider the following prescription for building up a galaxy distribution.
Throw down a galaxy at random. Now with some fixed probability, $p$, put another galaxy at a distance $L$ (in any direction) from it. Using the new galaxy as starting point repeat this process until you have the desired number of galaxies. Now there is no regular grid of galaxies but $L$ is still a preferred length in the distribution of the galaxies, and forms an SSR. To reconstruct this preferred scale is a statistical problem. This is illustrated schematically in Figure~(\ref{rings:fig}), which shows many rings of the same characteristic radius $L$ superimposed on one another. This superposition of rings on the plane visually `hides' the characteristic scale, as number of rings increases, and the sampling of each individual ring is reduced.

BAO, as we discuss below, provide an elegant SSR hidden between the rest of the galaxy clustering, but they are not the only possible SSR's. Any preferred scale in the clustering provides either an opportunity to apply a relative (absolute) Alcock-Paczynski test \cite{aptest} in the case the we don't (do) know {\em a priori} the size of the preferred scale, $L$. Other preferred scales include the Hubble scale at matter-radiation equality (which controls the scale of the turnover in the matter power spectrum) and the Silk damping scale. There may well be other preferred scales imprinted into the primordial clustering of matter. These can be naturally achieved if one inserts a short period of fast rolling into the otherwise slow-roll of inflation. The sudden change of inflaton velocity creates a bump in the matter power spectrum that can serve the same purpose as the BAO. The required fast-roll can be achieved in multi-field models of inflation.
\begin{figure}[htbp!]
\begin{center}
\includegraphics[width = 3.8in, height = 3.1in]{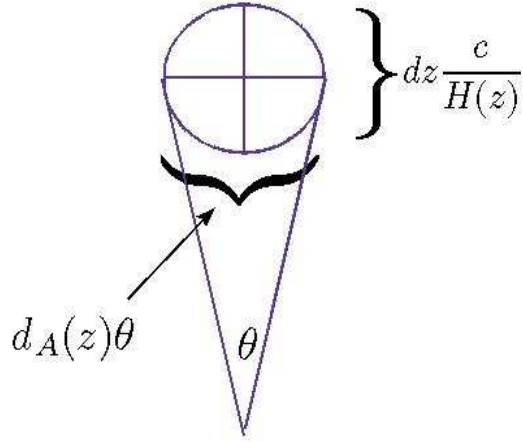}
\caption{The radial length of an object is given by $c dz/H(z)$ where $dz$ is the difference in redshift between the front and back of the object while the transverse size of the object is $d_A(z) \theta$ and $\theta$ is its angular size. If one knows that the object is spherical (but does not know the actual diameter) then one has the Alcock-Paczynski test which gives the product $d_A(z) H(z)$ from measuring $dz/\theta$. If, as in the case of BAO, one can theoretically determine the diameter, one has the bonus of finding $d_A(z)$ and $H(z)$ separately. \label{ap_schematic}}
\end{center}
\end{figure}

The SSR provided by the BAO has an additional advantage: it is primarily a linear physics phenomenon, which means we can ignore nonlinear effects\footnote{These nonlinear effects include redshift space distortions and nonlinear gravitational clustering, which will in general change the spherical nature of the oscillation scale.} to good approximation (we will discuss them later however). This also means we can turn the BAO into a calibrated or absolute Alcock-Paczynski test since the characteristic scale of the BAO is set by the sound horizon at decoupling. As a result the angular diameter distance and Hubble rate can be obtained separately. The characteristic scale, $s_{||}(z)$, along the line-of-sight provides a measurement of the Hubble parameter through \be H(z) = \frac{c\Delta z}{s_{||}(z)}\label{pll}\,,\ee while the tangential mode provides a measurement of the angular diameter distance, \be d_A(z)  = \frac{s_{\perp}}{\Delta \theta (1+z)}\label{perp}\,.\ee This is illustrated in the schematic Figure.~(\ref{ap_schematic}). The horizontal axis is Eq.~(\ref{pll}) and the vertical axis is $c\Delta z/H(z),$ Eq.~(\ref{perp}).

\begin{figure}[htbp!]
\begin{center}
\includegraphics[width = 5in]{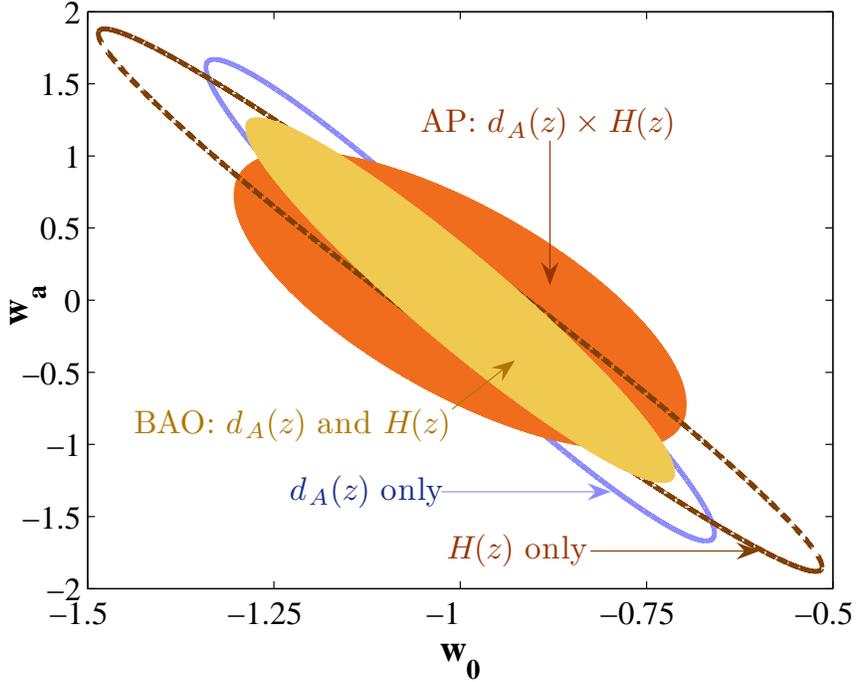}
  \caption{Comparing the Alcock-Paczynski (AP) test and BAO constraints -- on the dark energy parameters $w_0, w_a$ in the CPL parameterisation. The solid blue and dashed dark brown lines ellipses illustrate the constraints from a pair of $1\%$ measurements at $z=1,3$ of the angular diameter distance and Hubble parameter respectively. The filled beige ellipse shows the constraints from a BAO-like survey, where both the Hubble rate and angular diameter distance are measured simultaneously at $z=1,3$ (the errors on each have been increased by $\sqrt{2}$). In contrast, the dotted line illustrates constraints from the AP test: where $1.5\%$ measurements are made on the product $H\times d_A$ at the same redshifts. The AP test provides comparable constraints to using only a single measurement while the BAO provide the tightest constraints; they measure $d_A$ and $H$ simultaneously. The assumed fiducial model is $(H_0, \Omega_m, \Omega_k, w_0, w_a) = (70, 0.3, 0, -1,0),$ and the priors on the model are given as $\mathrm{Prior}(H_0, \Omega_m, \Omega_k, w_0, w_a) = (10^4, 10^4, 10^4, 0, 0).$ Figure produced using \name~\cite{fisher_release}.  \label{aptest:fig}}
\end{center}
\end{figure}

While the AP test on its own constrains the product $d_A(z) \times H(z),$ it is just one function, and so combining the measurements of $d_A$ and $H$ through the BAO provide tighter constraints on cosmological parameters. This is illustrated in Figure~(\ref{aptest:fig}), which shows the error ellipse in the dark energy parameters from a hypothetical galaxy redshift survey with constraints from the angular diameter distance and Hubble parameter and the product $H \times d_A$ from the AP test. Constraints on the dark energy parameters from the AP test are similar to those from a single observable such as $d_A$, while the constraints are significantly improved when combining measurements of both $H$ and $d_A.$

 One method of extracting a statistical scale from the clustering of galaxies is via the two-point correlation function, $\xi(r),$ which quantifies the excess clustering on a given scale relative to a uniform distribution with the same mean density. The correlation function of galaxies is approximately described by a power law \cite{totsuji}, \be \xi(r) \propto \left(\frac{r_0}{r}\right)^{\gamma}\,,\ee with $r_0 \sim 5 h^{-1}\mathrm{Mpc}^{-1}.$

\begin{figure}[htbp!]
\begin{center}
$\begin{array}{@{\hspace{-0.2in}}c@{\hspace{-0.22in}}c}
\epsfxsize=3in
\epsffile{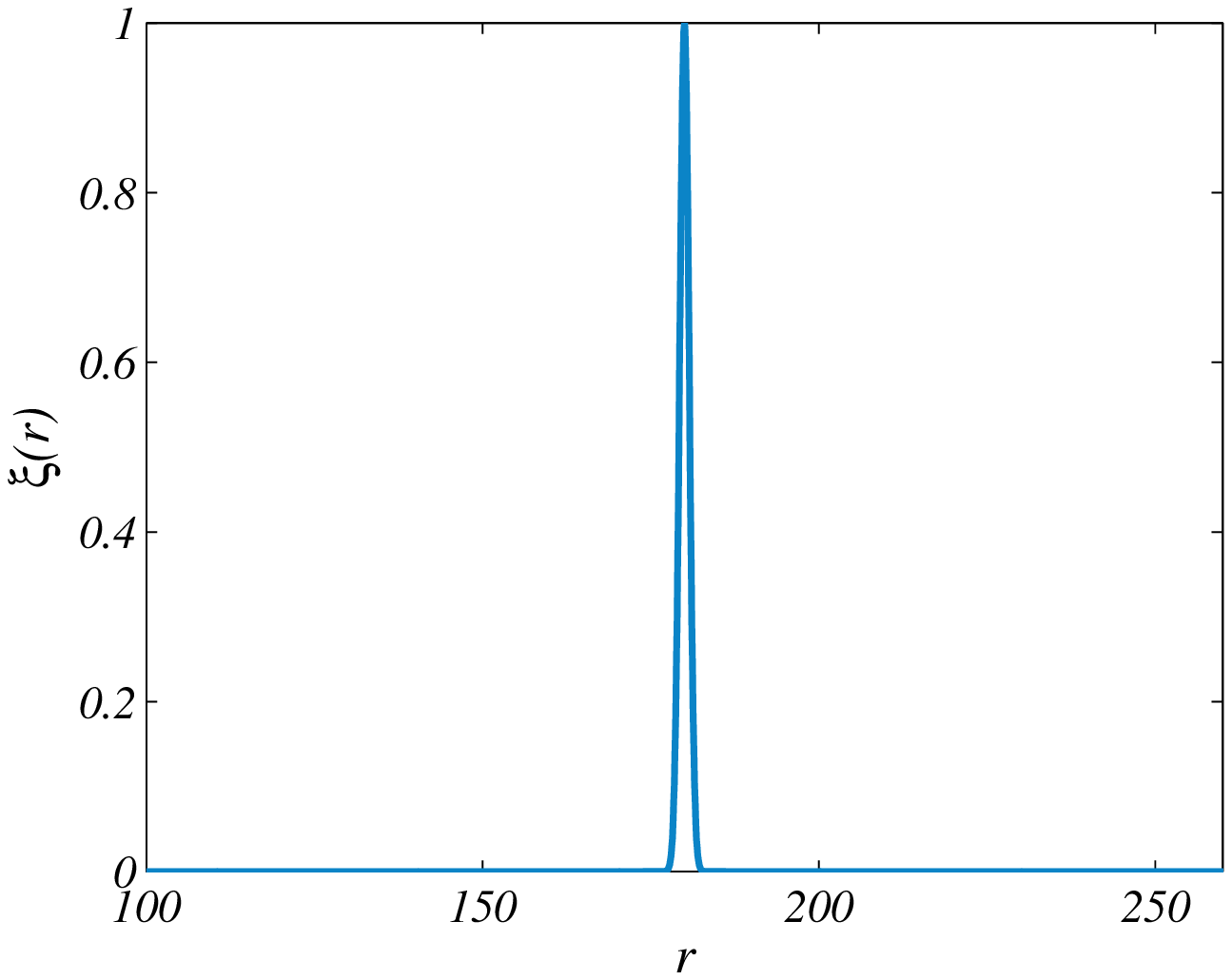} &
	\epsfxsize=3in
	\epsffile{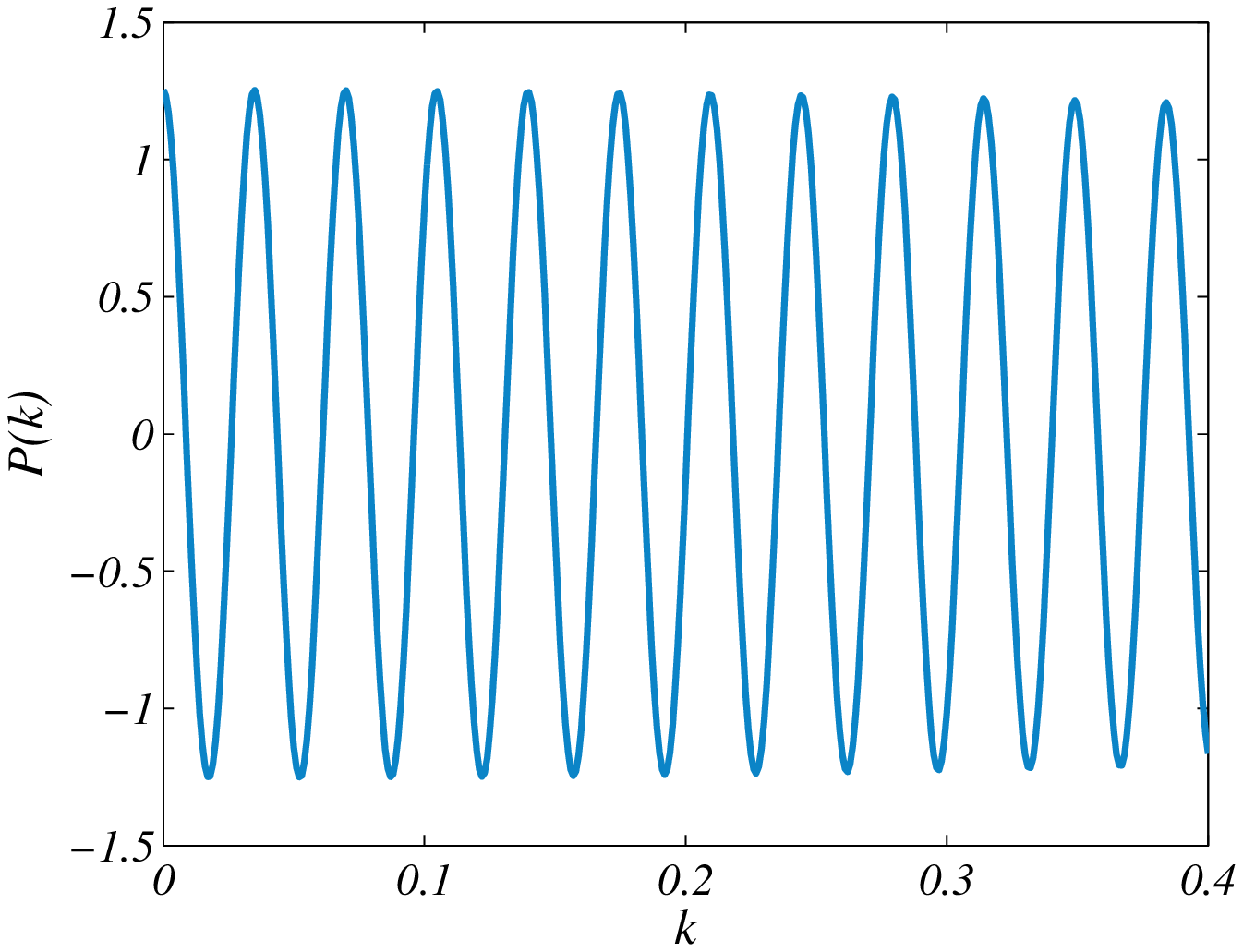}\\ [0.0cm]
 \end{array}$
  \caption{Schematic illustration of the Fourier pairs $\xi(r), P(k)$. A sharp peak in the correlation function (left panel) corresponds to a series of oscillations in $P(k)$ (right panel). The Baryon Acoustic Peak in the correlation function will induce characteristic \bao~in the power spectrum. \label{schematic:fourier}}
  \end{center}
 \end{figure}

A characteristic scale in the clustering of galaxies will appear as a peak or dip in the correlation function, depending on whether there is an excess or deficiency of clustering at that scale. Any characteristic features will also be present in the power spectrum, since the correlation function and power spectrum (we consider for simplicity the simple 1-dimensional spherically averaged power spectrum) form a Fourier pair:  \be P(k) = \int_{-\infty}^{\infty} \xi(r)\exp(-ikr)r^2dr\,.\ee We will now see how features in the two functions are related.
A $\delta$ function at a characteristic scale, say $r_*,$ in $\xi(r)$ will result in power spectrum oscillations, $P(k) \propto e^{-ikr_*}, $ as can be seen in Figure~(\ref{schematic:fourier}). These are the \bao.
 \\
\begin{figure}[htbp!]
\begin{center}
\includegraphics[width = 3.8in]{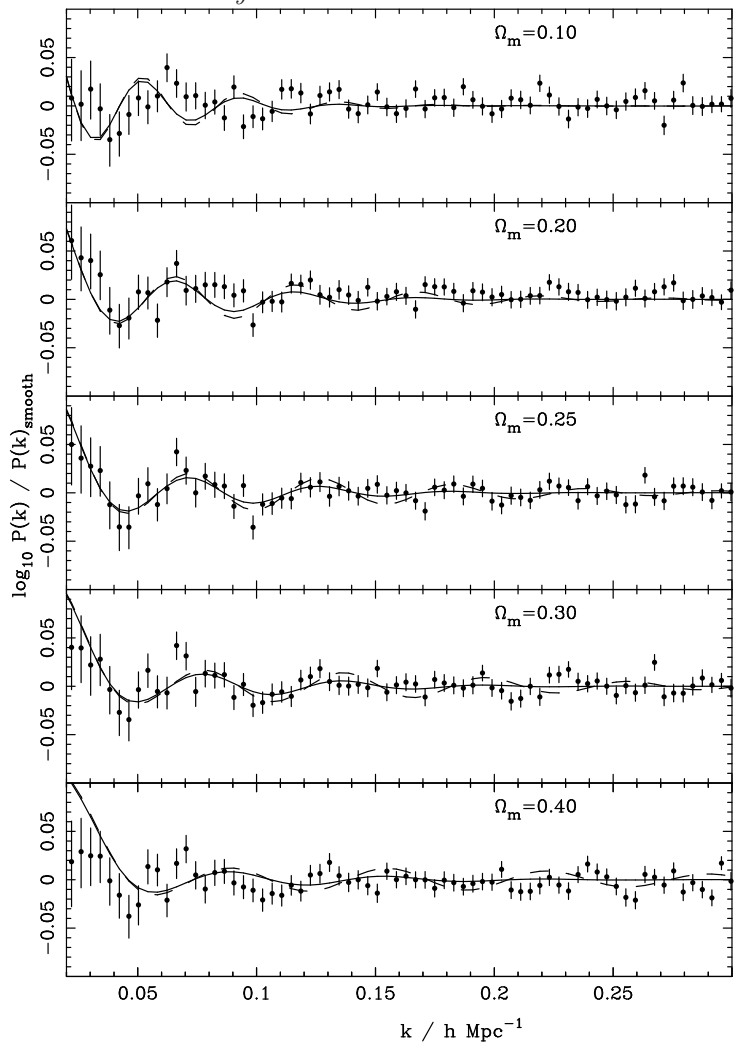}
\caption{Changing cosmology moves the \bao~-- the ``wiggles-only'' power spectra of the SDSS LRG survey. In each panel, a fiducial model with a different value of $\Omega_m$ has been used to convert the data from redshift to comoving distance, from $\om = 0.1$ (top) to $\om = 0.4$ (bottom). The solid lines in each panel show the CDM prediction for the BAO assuming the particular value of $\om$, while the dashed lines show the same model without the low-redshift small-scale damping term. In each case the baryon fraction is held fixed at $17\%$, and $h = 0.73$. As the matter density is changed, both the theoretical scale of the BAO changes, and the data moves around through the conversion from redshift to comoving distance. Figure from Percival {\em et al.,} 2006 \cite{percival_lrg}. \label{percival_lrg:fig}}
\end{center}
\end{figure}
These characteristic oscillations are powerful probes of dark energy.
If we used only the radial Fourier modes we would obtain a measurement
of $H(z)$ while the purely transverse modes yield a pure measurement
of $d_A(z)$. If we limited ourselves to just the purely radial and
transverse modes we would be throwing away a large number of modes
(and hence information) since most Fourier modes are not purely radial
or transverse of course, but rather contain components of both. As a
result, the measurements of $H(z)$ and $d_A(z)$ from actual BAO
surveys are actually partially anti-correlated and hence not
independent. This anti-correlation actually leads to stronger
cosmological constraints than an uncorrelated analysis would suggest, as described in \cite{seo_eis07}.

A nice example of their sensitivity and the issues involved in using BAO for parameter estimation is evident in Fig.~(\ref{percival_lrg:fig}) from Percival {\em et al.} \cite{percival_lrg} which shows both the SDSS LRG data and theoretical fits as a function of $\Omega_m$. The panels each show
the residual BAO oscillations relative to a non-oscillatory reference spectrum (a technique advanced in Blake {\em et al.} \cite{blake_glaze_simulation}), thus isolating the BAO information alone. The panels have a fixed baryon fraction of $17\%$ relative to dark matter. An important point to take away from this plot is that both the theory {\em and the data} vary with $\Omega_m$.

This situation is different from more standard model fitting exercises where the data is static as the model parameters are varied. Here, to compute the data points, a redshift-distance relation
(which depends on $\Omega_m$ of course) must be assumed to move the points into physical, as opposed to redshift space, after which the power spectrum is constructed. In \cite{percival_lrg}, 31 data sets and theoretical predictions were computed for various flat $\Lambda$CDM models
with different values of $\Omega_m$. The impact of varying $\Omega_m$ is seen mostly clearly on the theoretical curves where the amplitude and frequency of oscillation is affected, both by
the change to the underlying BAO scale (since the $\rho_b/\rho_{\gamma}$
ratio changes and hence the speed of sound prior to decoupling) and
the redshift-distance relation which will move galaxies in the radial direction, bringing them either closer or moving them further away in distance.
This dependence of the data on the assumed cosmology is typically
ignored in parameter estimation studies using
the BAO results, an approximation that is fairly good if one does not
range far from the
fiducial model \cite{tegmark_lrg, percival_0705.3323}. The dilation resulting from using a different fiducial model would scale the $k$-axis of the dimensionless power spectrum by the ratio \cite{eisenstein_05,tegmark_lrg}: \begin{equation}
a = \frac{d_V(z)}{d_V^{\mathrm{fiducial}}(z)},
\end{equation} where $d_V(z) \equiv \( d_A(z)^2cz/H(z)\)^{1/3} $ combines the radial and transverse dilation.

Tegmark {\em et al.} calibrate this for the parameter range allowed by the combined SDSS-LRG and WMAP3 data, and find that corrections to the $k$-scale are  typically less than or equal to $3\%.$ As an example, using a fiducial model with $\om = 0.25$ results in a bias of the measured value of $\om$ of $\sim2\%.$
\subsection{Physics of the BAO \label{linphys}}
\begin{figure}[htbp!]
\centerline{\epsfxsize=2.9in\epsffile{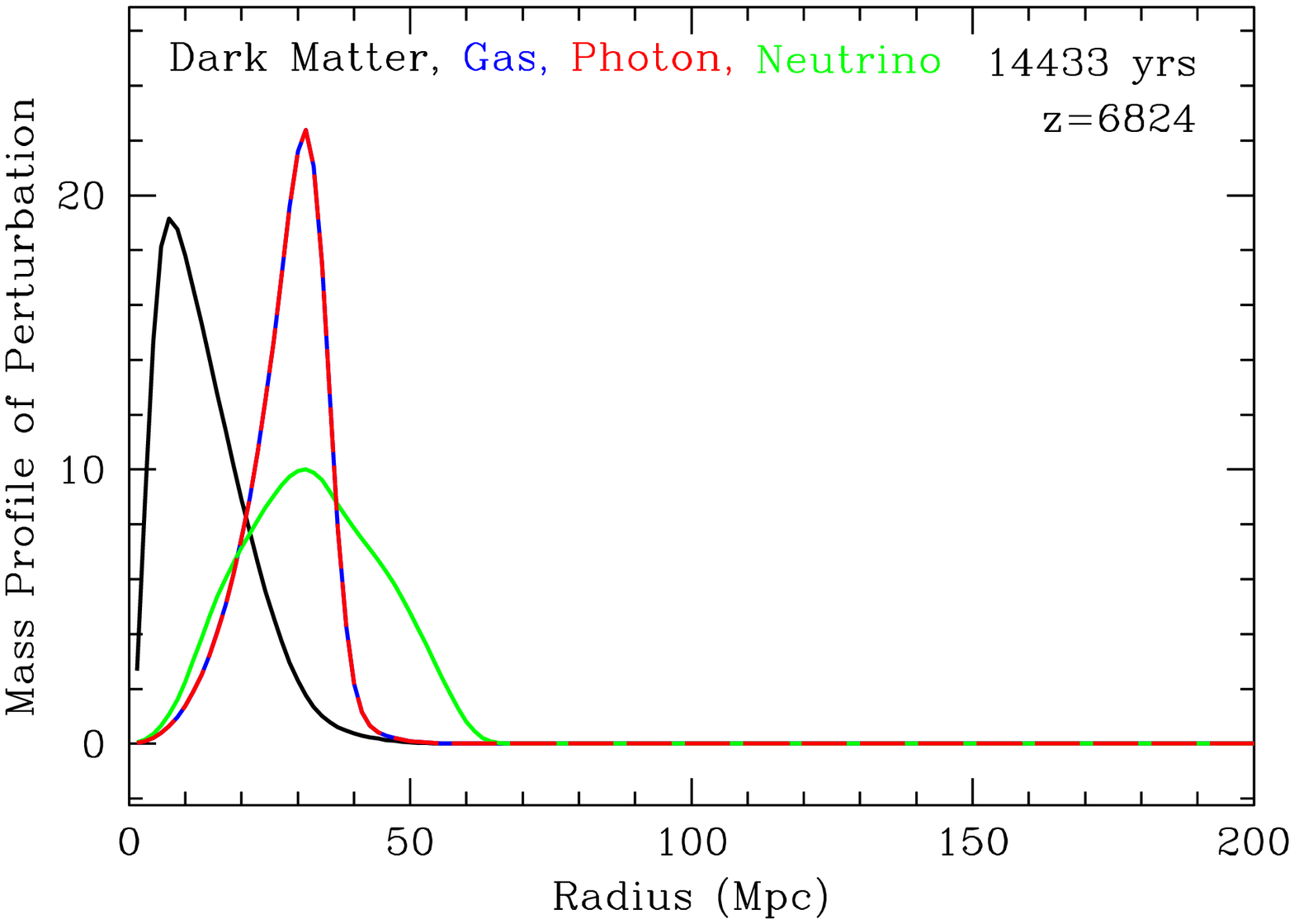}
    \quad   \epsfxsize=2.9in\epsffile{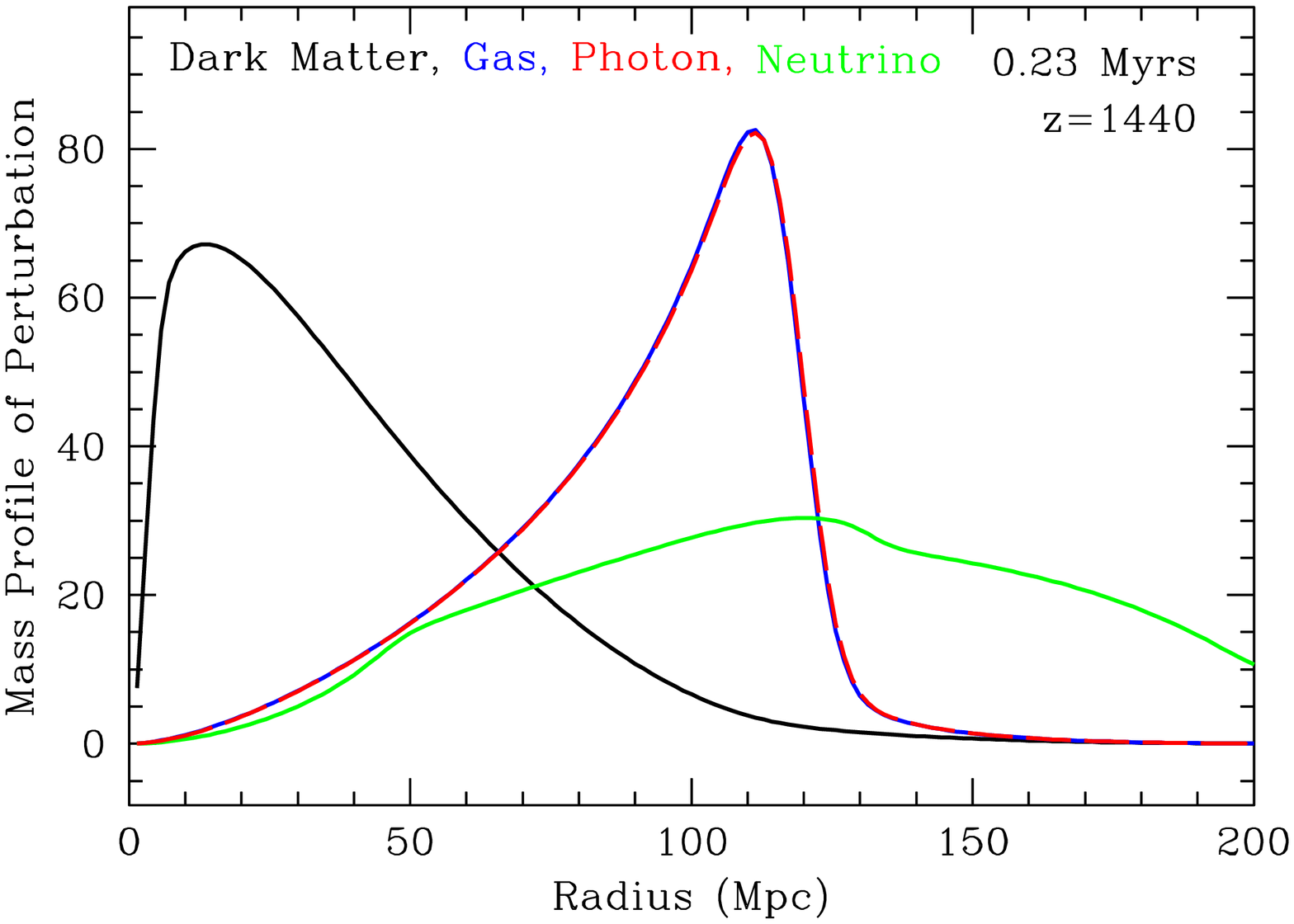}}
\centerline{\epsfxsize=2.9in\epsffile{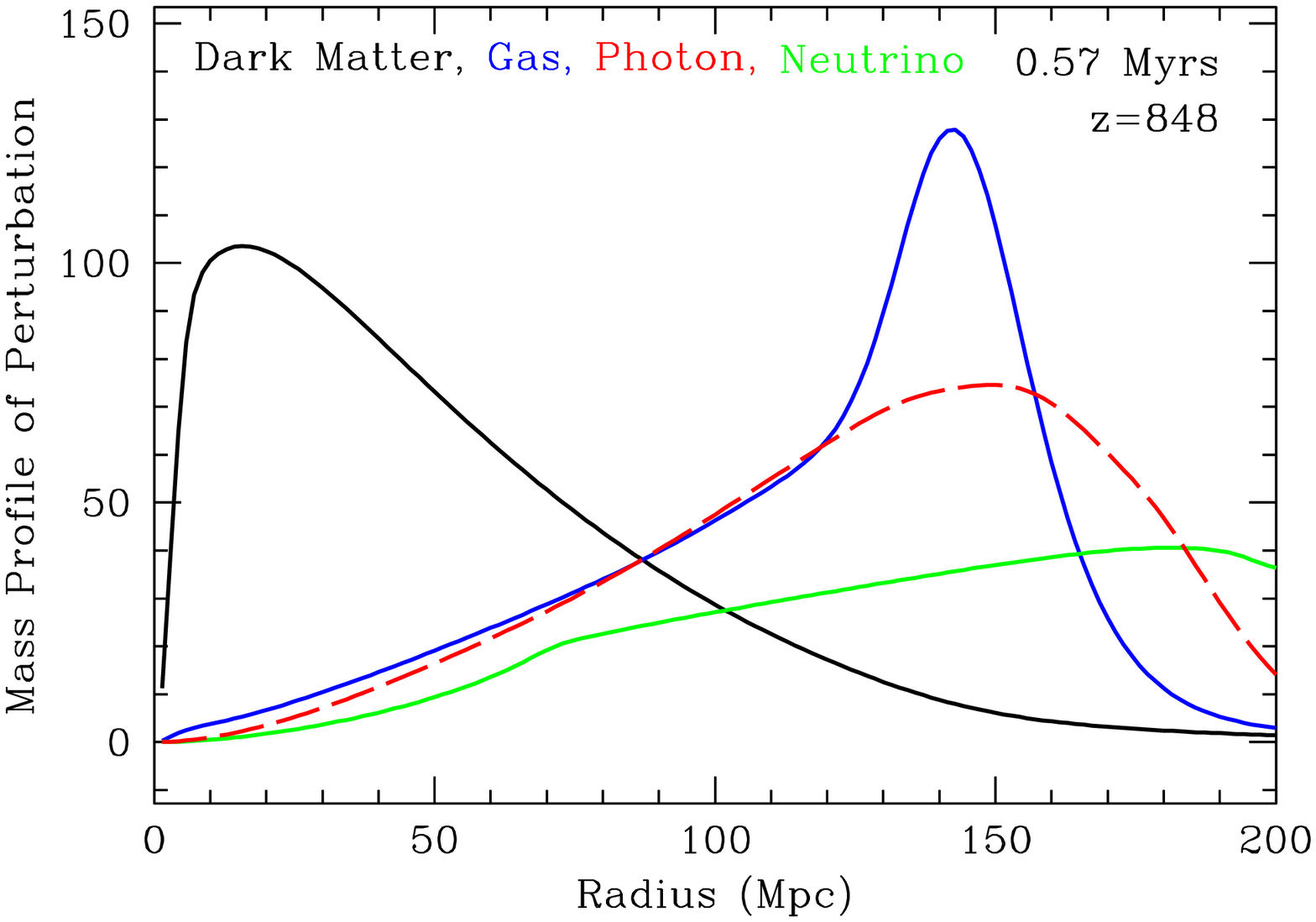}
    \quad   \epsfxsize=2.9in\epsffile{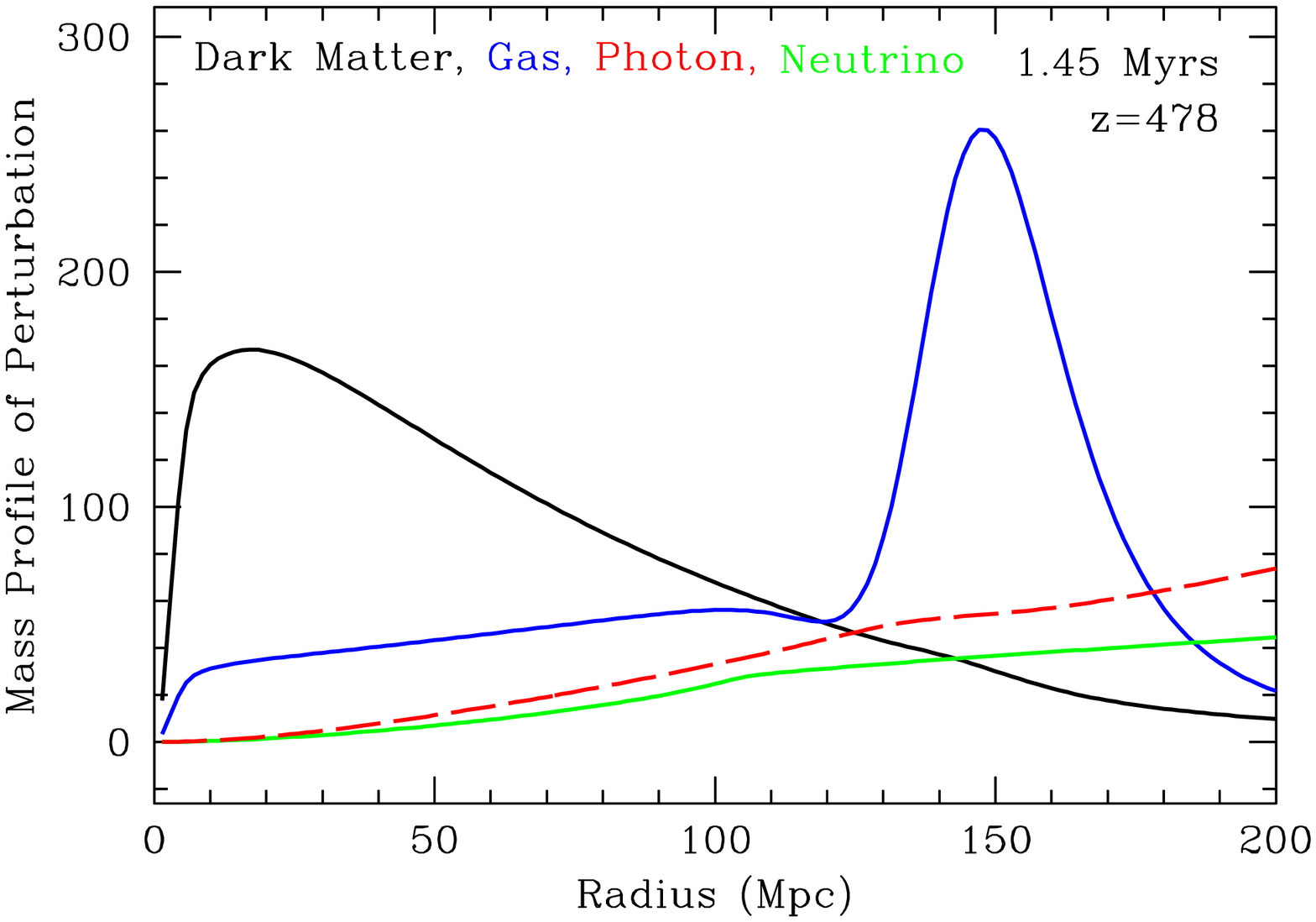}}
\centerline{\epsfxsize=2.9in\epsffile{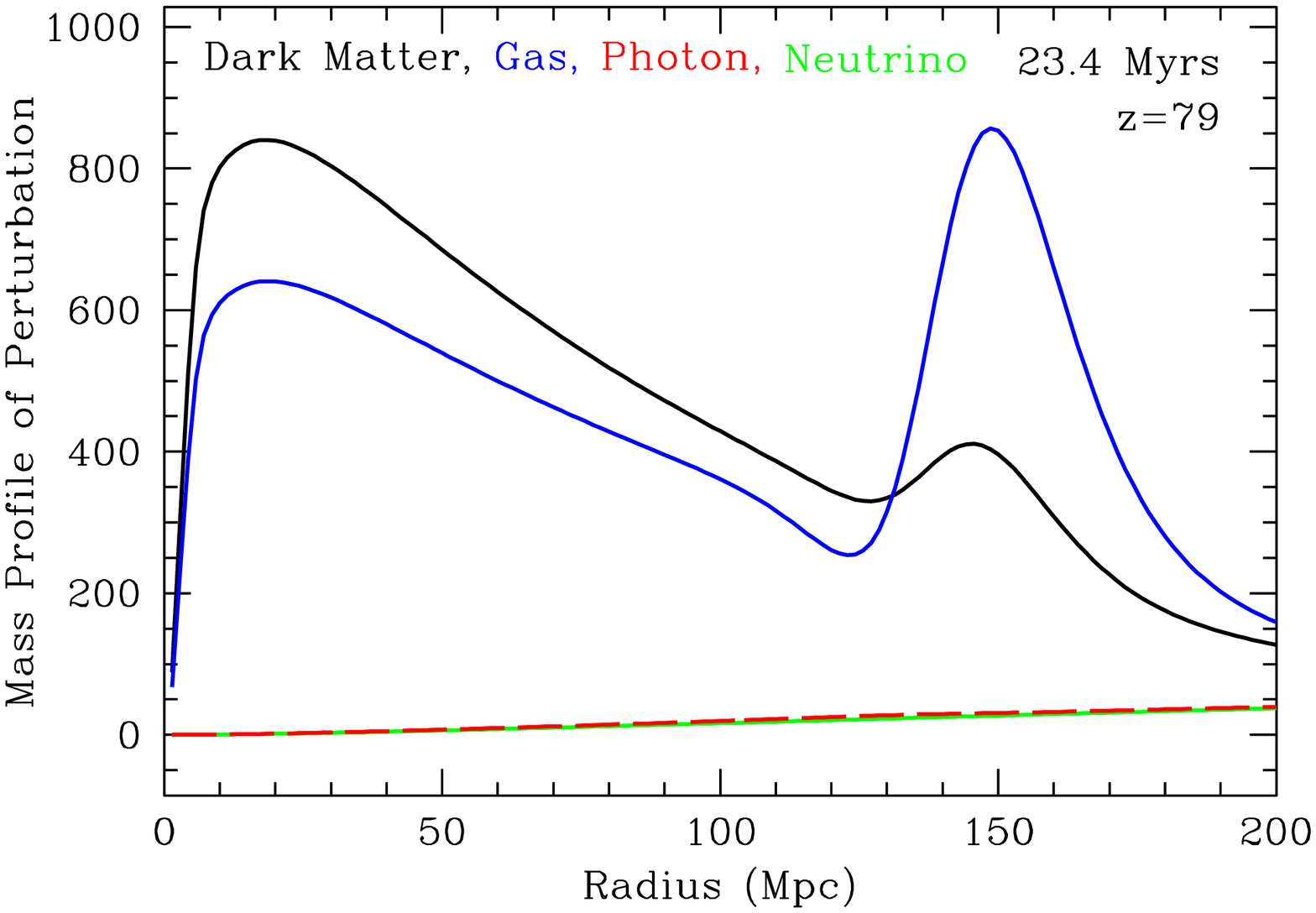}
    \quad   \epsfxsize=2.9in\epsffile{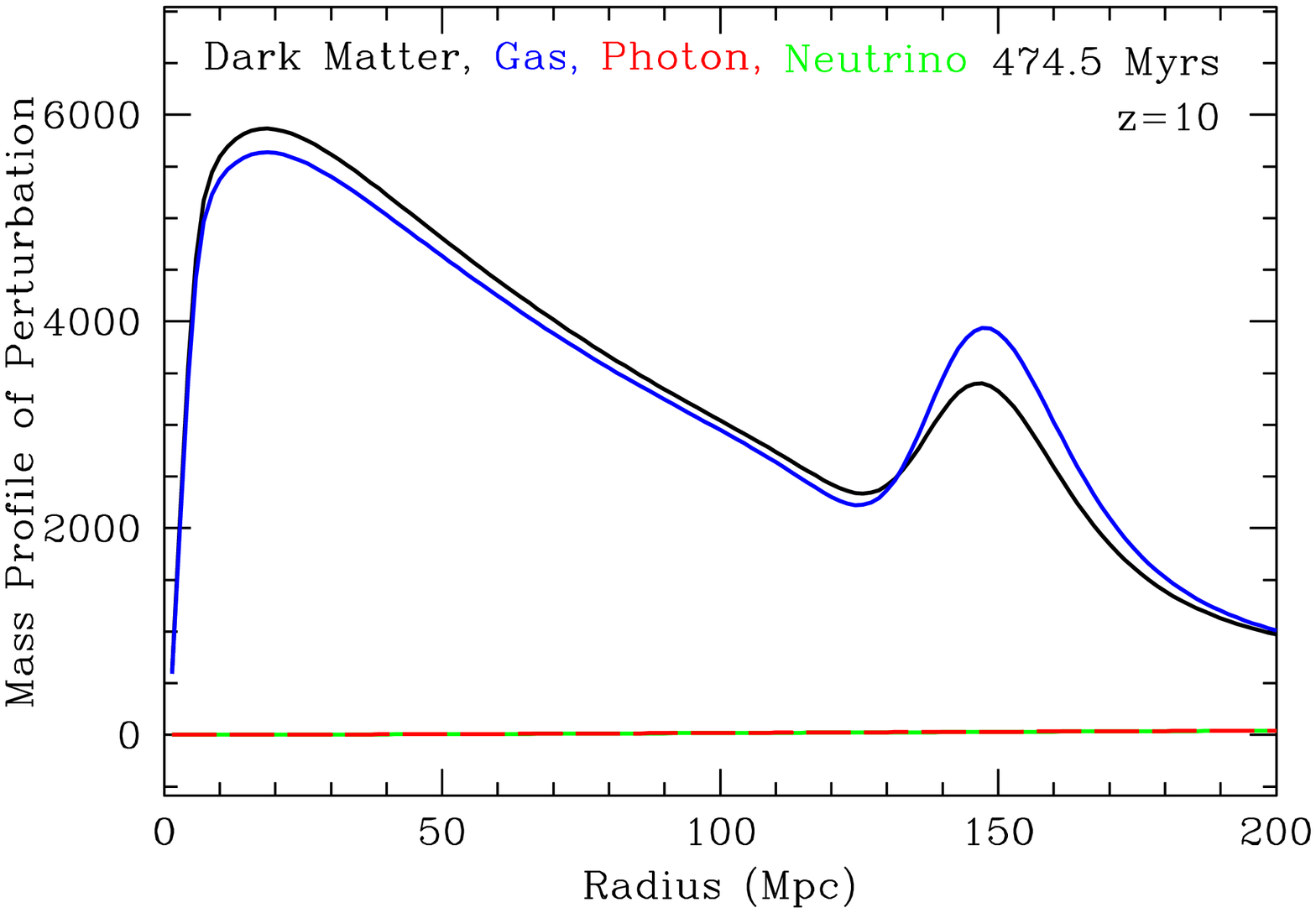}}
\caption{{\bf Snapshots of an evolving spherical density perturbation --} the radial mass profile as a function of comoving radius for an initially point-like overdensity located at the origin. The perturbations in the dark matter (black curve), baryons (blue curve), photons (red) and neutrinos (green) evolve from early times ($z=6824$, top left) to long after decoupling ($z=10$, bottom right). Initially the density perturbation propagates through the photons and baryons as a single pulse (top left-hand panel). The drag of the photons and baryons on the dark matter is visible in the top right panel; the dark matter only interacts gravitationally and therefore its perturbation lags behind that of the tightly coupled plasma. During recombination, however, the photons start to ``leak'' away from the baryons (middle left panel); and once recombination is complete $(z = 470$, middle right) the photons freely steam away leaving only a density perturbation in the baryons around $150\mathrm{Mpc},$ and a dark matter perturbation near the origin. In the bottom two panels we see the how the gravitational interaction between the dark matter and the baryons affects the peak: dark matter pulls the baryons to the peak in the density near zero radius, while the baryons continue to drag the dark matter overdensity towards the $150\mathrm{Mpc}$ peak (bottom left), finally yielding a peak in the radial mass profile of the dark matter at the scale set by the distance the baryon-photon acoustic wave could have travelled in the time before recoupling. Figure taken from Eisenstein {\em et al.} \cite{ESW}. \label{eisenstein_shell:fig}}
\end{figure}

Before recombination and decoupling the universe consisted of a hot plasma of photons and baryons which were tightly coupled via Thomson scattering. The competing forces of radiation pressure and gravity set up oscillations in the photon fluid. If we consider a single, spherical density perturbation in the tightly coupled baryon-photon plasma it will propagate outwards as an acoustic wave with a speed $c_s = c/\sqrt{3(1+R)},$ where $R \equiv 3\rho_b/4\rho_\gamma \propto \Omega_b/(1+z)$ \cite{ESW}. At recombination the cosmos becomes neutral and the pressure on the baryons is removed. The baryon wave stalls while the photons freely propagate away forming what we now observe as the Cosmic Microwave Background (CMB). The characteristic radius of the spherical shell formed when the baryon wave stalled is imprinted on the distribution of the baryons as a density excess. The baryons and dark matter interact though gravity, and so the dark matter also preferentially clumps on this scale. There is thus a increased probability that a galaxy will form somewhere in the higher density remains of the stalled baryon wave than either side of the shell. This scenario is illustrated in the iconic Figure~(\ref{eisenstein_shell:fig}) from Eisenstein {\em et al.} \cite{ESW}.

Had a galaxy formed at the centre of our initial density perturbation there would be a bump in the two-point correlation function at the radius $s$ of our spherical shell, reflecting the higher probability of finding two galaxies separated by the distance $s$. The scale $s$ is usually close to the sound horizon\footnote{This is perhaps a slight misnomer since $s$ is not the maximum distance that could have been travelled by any fluid. A simple scalar field has speed of sound equal to $c$ and hence travels faster than the baryon-photon sound wave we are discussing here. }, the comoving distance a sound wave could have travelled in the photon-baryon fluid by the time of decoupling, and depends on the baryon and matter densities via \cite{eis_white}:
\be
s = \int_{z_{rec}}^\infty\frac{c_sdz}{H(z)} = \frac{1}{\sqrt{\Omega_mH_0^2}}\frac{2c}{\sqrt{3z_{eq}R_{eq}}}\ln
\left[\frac{\sqrt{1+R_{rec}}+ \sqrt{R_{rec}+R_{eq}}}{1+\sqrt{R_{eq}}}\right]\,,
\ee
where $R \equiv 3\rho_b/4\rho_\gamma \propto \Omega_bh^2/(1+z), z_{eq} = \Omega_m/\Omega_{rad}$ is the redshift of matter-radiation equality and ``rec'' refers to recombination. The CMB strongly constrains the matter and baryon densities at decoupling and hence the sound horizon, $146.8 \pm1.8 \mathrm{Mpc}$ \cite{wmap5}\footnote{Confusion sometimes arises from quoting the BAO scale as both $\sim 150 \mathrm{Mpc}$ and $105 h^{-1}\mathrm{Mpc}$. Here $h\simeq 0.7$ is the Hubble constant today in units of $100\mathrm{km/s/Mpc}$ and provides the required conversion factor.}. Hence this scale is itself an excellent standard ruler as long as one can measure $\Omega_b$ to high precision, and in the case where one allows for exotic radiation components in the early Universe, the redshift of equality \cite{eis_white}. Of course, the early universe was permeated by many such spherical acoustic waves and hence the final density distribution is a linear superposition of the small-amplitude sound waves in the usual Green's function sense. Indeed this real space approach to cosmic perturbations based on Green's functions can be made rigorous, see Bashinsky and Bertschinger \cite{bert}.
\section{Forecasting BAO Constraints and Power Spectrum Errors}
To make projections for the ability of a BAO survey to constrain cosmological parameters, one can either simulate data given survey parameters or perform a Fisher matrix analysis, \cite{tegmark,seo2003,matsubara,SE2005,amendola_de,seo_eis07,amendola_growth}. The Fisher matrix, $F_{ij},$ defines how survey accuracy on physical observables (for e.g. the power spectrum or correlation function) translates into constraints on parameters of interest, $\theta_i$, such as curvature or dark energy, with the marginalised error on the $i$-th parameter of interest being given via the $ii$-component of the inverse of the Fisher matrix:
\be
\Delta \theta_i \geq \sqrt{(F^{-1})_{ii}}
\ee
while in the case where one does not marginalise over any other parameters the best error is simply given by $\Delta \theta_i \geq (F_{ii})^{-1/2}$ which is always smaller. In the case where the likelihood is exactly Gaussian in the parameters then these error estimates are exact and not simply lower-bounds.

In the case of the \bao~the Fisher matrix can be given directly in terms of derivatives of the power spectrum $P(\overrightarrow{k})$ with respect to the cosmological parameters of interest, $\theta_i$ \cite{seo2003} as a function of the vector $k$:
\bea F_{ij} &=& \int_{\overrightarrow{k}_{\mathrm{min}}}^{\overrightarrow{k}_{\mathrm{max}}} \frac{\partial \ln P(\overrightarrow{k})}{\partial \theta_i}\frac{\partial \ln P(\overrightarrow{k})}{\partial \theta_j}V_{\mathrm{eff}}(\overrightarrow{k}) \frac{d^3\overrightarrow{k}}{2(2\pi)^3} \nonumber \\
&=& \int_{-1}^{1}\int_{k_{\mathrm{min}}}^{k_{\mathrm{max}}} \frac{\partial \ln P(k,\mu)}{\partial \theta_i}\frac{\partial \ln P(k, \mu)}{\partial \theta_j}V_{\mathrm{eff}}(k, \mu) \frac{2\pi k^2 dk d \mu}{2(2\pi)^3}\,. \label{powerspec}\eea
Assuming azimuthal symmetry for $P$ along the line of sight means that $P(\overrightarrow{k})$ can be described by $\mu = \overrightarrow{k}\cdot \hat{r}/k$, with $\hat{r}$ is the unit vector along the line of sight and $k = |\overrightarrow{k}|.$ Eq.~(\ref{powerspec}) is valid  between wavenumbers between $k_{\mathrm{min}}$ and a maximum wavenumber to exclude the non-linear regime. A conservative value for $k_{\mathrm{max}}$ is $ 0.1~h\mathrm{Mpc}^{-1}$ at $z=0$ \cite{eisenstein_05}.

A crucial ingredient for the Fisher matrix above is the effective volume: \bea V_{\mathrm{eff}}(k,\mu) &=& \int \left(\frac{n(\overrightarrow{r})P(k,\mu)}{n(\overrightarrow{r})P(k, \mu) + 1}\right)^2d^3\overrightarrow{r} \nonumber \\
&=&\left(\frac{nP(k, \mu)}{nP(k, \mu) +1}\right)^2V_{\mathrm{survey}}, \eea where the final expression assumes that the number density $n(\overrightarrow{r})$ is independent of position. The effective volume transforms the theoretical, survey-independent derivatives into specific predictions for the survey one is considering.  The effective volume encodes the accuracy with which the survey under consideration can measure the power spectrum at different wavenumbers, $k$. To gain insight into this key element, let us model the error on the power spectrum as \cite{tegmark,blake_bao}, \be \frac{\delta P}{P}= \frac{1}{\sqrt{m}}\left(1+ \frac{1}{nP}\right)\,,\label{accuracy}\ee where $m$ is the total number of independent Fourier modes contributing to the measurement of the oscillation scale, and $P \equiv P(k^*), ~k^* \simeq 0.2 h \mathrm{Mpc}^{-1}$ is the value of the power spectrum amplitude at an average scale $k^*$,  characteristic of the acoustic oscillations. From Eq.~(\ref{accuracy}) we note that the two competing sources of error in reconstructing the baryon acoustic oscillation scale are cosmic variance and shot noise, represented by the first and second terms respectively.
\subsection{Shot noise}
\begin{figure}[htbp!]
\begin{center}
\includegraphics[width = 5.8in, height = 4.9in]{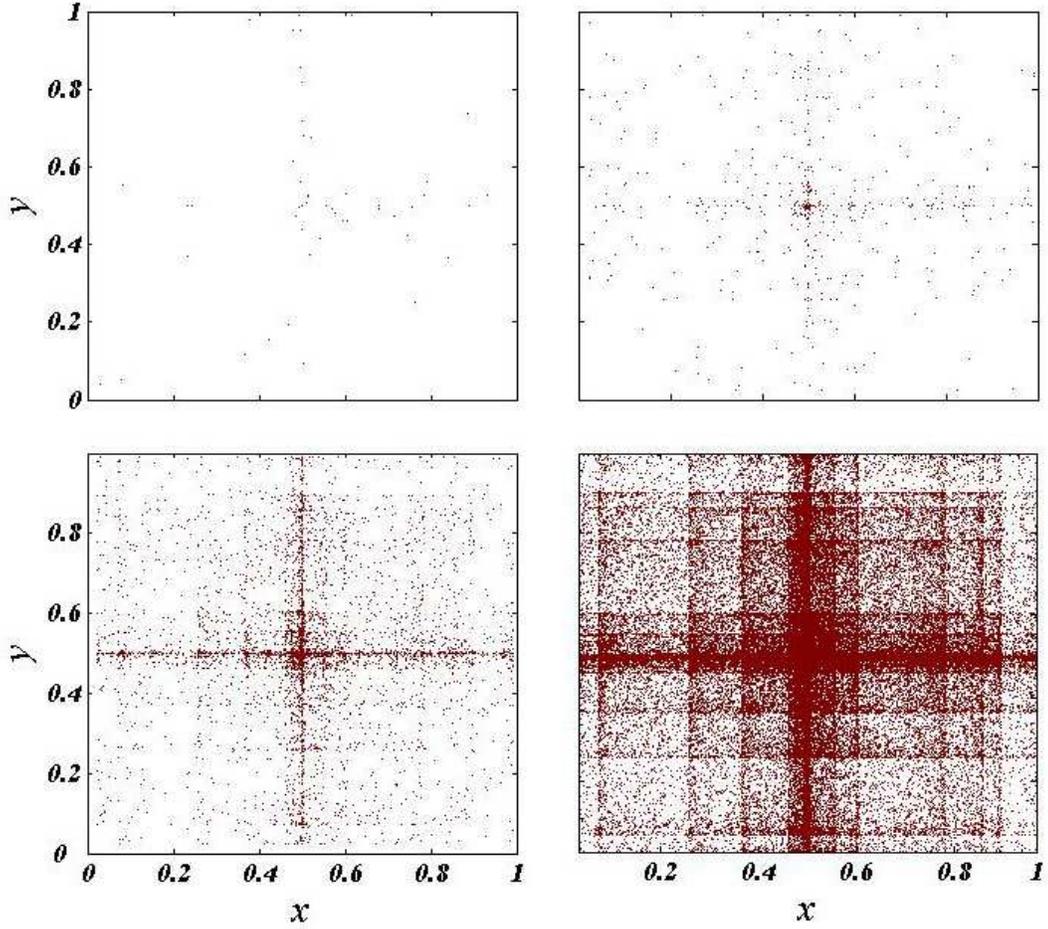}
\caption{The effect of {\bf shot noise} -- as the number of galaxies in a survey increases, one is able to reconstruct the underlying pattern in the distribution of those points more reliably. This is illustrated in the progression from the top left hand panel {\bf(100 points)} to the bottom right hand panel ({\bf 100 000 points}) which are all drawn from the same probability distribution. As the number of points increases, the sub-structure of the pattern becomes visible. \label{shot_noise}}
\end{center}
\end{figure}

Our goal is to reconstruct the underlying dark matter distribution from discrete tracers such as galaxies. As is illustrated schematically in Figure~(\ref{shot_noise}), this reconstruction is very difficult if we have few objects, an effect known as Poisson shot noise. Shot noise error decreases as the density of targets in a given volume increases: the complex underlying pattern in Figure~(\ref{shot_noise}) becomes clearer as the number of points increases and the pattern is sufficiently sampled. Increasing the number of targets requires longer integration times on the same patch of sky to go deeper, leading to a reduction in the area surveyed in a fixed observing time.

Recently, an exciting proposal has been made to drastically reduce the shot noise term \cite{seljak_shotnoise}. In the context of the halo model, the power spectrum of dark matter is made up by summing the contributions from dark matter halos of different sizes. Under the assumption that galaxies form within the halos according to a given halo occupation distribution, one can derive a similar expression for the power spectrum of galaxies. The central idea in \cite{seljak_shotnoise} is that nonlinear evolution of structure in the dark matter ensures that the power spectrum of dark matter behaves as $k^4$ for $k\rightarrow 0$, as opposed to the usual constant ($k^0$) Poisson term. Using an N-body simulation with a scheme where central halo galaxies are weighted by halo mass, Seljak \emph{et al.} \cite{seljak_shotnoise} find that the shot noise term is suppressed by a factor $10-30$.
\subsection{Cosmic variance}
\begin{figure*}[htbp!]
  \begin{flushleft}
$\begin{array}{@{\hspace{-0.25in}}c}
\includegraphics[width = 6in]{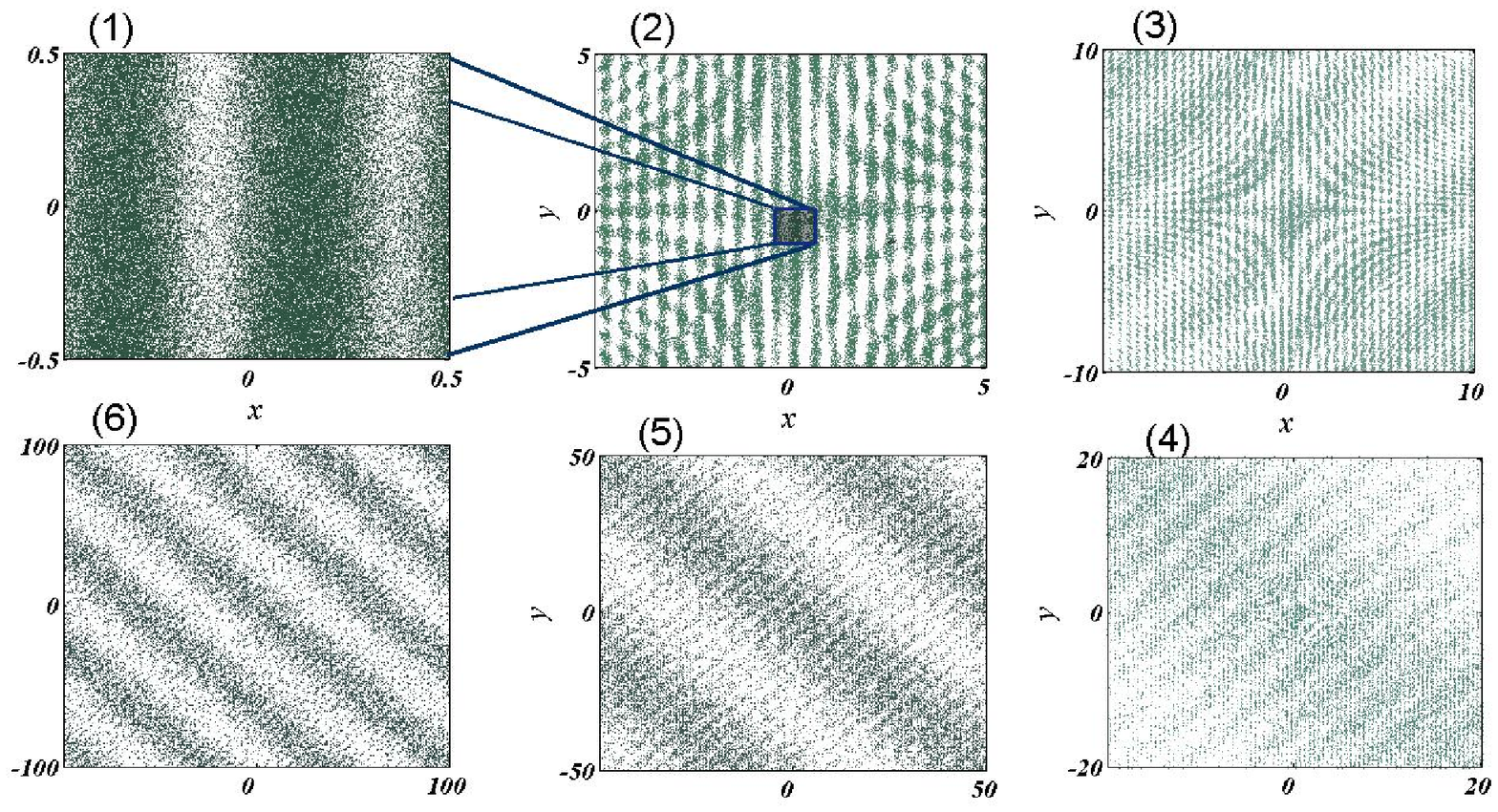}
\end{array}$
    \end{flushleft}
    \begin{center}
  \caption{Illustrating {\bf cosmic variance} with successive zooms -- as the window expands (clockwise from top left), so large-scale modes in the distribution of points become visible, due to the reduction in cosmic variance. The number of points is the same in each panel while the volume increases by a factor of $4 \times 10^4$ over the six panels. \label{cv}}
  \end{center}
 \end{figure*}
Cosmic variance is the error arising when we can't see the big picture: we cannot estimate the clustering on scales larger than our survey size. This is illustrated schematically in Figure~(\ref{cv}), where points are distributed according to a pattern consisting of Fourier modes with a variety of wavelengths and directions. The survey size increases in a clockwise direction, starting from top left, allowing the large-wavelength modes to become visible. To reconstruct these large-scale patterns requires an increased survey volume. In cosmology, it is impossible to keep increasing the size of the survey indefinitely as we are limited to the observable universe and so the values of the power spectrum on the very largest scales are fundamentally limited by this cosmic variance.
To minimise the cosmic variance requires sampling the largest possible volume, which at fixed observing time implies spending as little time integrating on each field as possible, i.e. the exact opposite of the optimal strategy to minimise shot noise. The total error on the power spectrum is therefore a combination of the effects of the finite size of the survey and the number density of objects used to sample the underlying distribution.

An interesting recent development is the suggestion by McDonald and Seljak \cite{mcdonald_seljak} that by using multiple tracers of the underlying dark matter distribution, with different biases, one can constrain the redshift distortion parameter $\beta \equiv b^{-1} d\ln G/d\ln a,$ to a level of precision controlled only by the shot noise, with the exciting possibility that cosmic variance may not be the limiting factor for certain surveys.

\subsection{Redshift Errors \label{sec:photz}}
To compute $P(k)$, or $\xi(r)$ we need the redshifts of the objects in our survey. This can be achieved by taking a spectrum, which typically gives a highly accurate redshift, $\delta z < 0.1\%,$ or by using the colours of the object alone to get a photometric redshift, which typically gives projections of $\delta z \sim 3-5\%,$ for 5 optical bands. The trade-off when going from a spectroscopic to a purely photometric survey is to sacrifice redshift accuracy for greater depth, area and volume. How does degrading redshift affect
the accuracy of the survey? Well, consider a fractional error
$\sigma_z$ on the redshift $z$. From our discussion of the radial
comoving distance and the Alcock-Paczynski effect, we know
that an error $\delta z$ will result in an uncertainty in the radial position of $\delta L = c \delta z/H(z)$. If we consider $z \sim 1$ we see that even if $\delta z/(1+z) = 0.01$, the radial uncertainty is
significant at around $1\%$ of the Hubble scale or about $40 h^{-1}\mathrm{Mpc}$. This is a large error when we are trying to measure the BAO on a scale of $\sim105 h^{-1}\mathrm{Mpc}$. It is significantly larger than the systematic errors due to nonlinear effects and worse we cannot calibrate for it, as will probably be possible for the nonlinear effects. In addition one must include the fact that photometric redshift errors are typically non-Gaussian. They usually exhibit catastrophic wings at certain redshifts which reflect confusion with objects with similar colours at completely different redshifts, which means the likelihood of getting a ``$5\sigma$" error is much larger than the Gaussian approximation would suggest.

However, these redshift errors only affect our knowledge of the radial
position of the galaxy. In the angular direction, astrometry errors ($\delta \theta$)
induce transverse distance errors of order $\delta L \sim \delta
\theta d_A(z)$. If we demand that this be less than $1 h^{-1} \mathrm{Mpc}$ at all
redshifts of potential interest ($z < 4$) we need astrometry accuracy
better than about $0.5'$ which is easily satisfied in all modern
galaxy surveys. Hence even if our knowledge of the radial
distribution of galaxies is significantly degraded in photometric
surveys, the angular distribution is preserved. As a result
measurements of the Hubble rate (derived from the radial BAO scale) in
photo-z surveys are more strongly affected than measurements of $d_A$
(derived from the transverse BAO scale).

\begin{figure}[htbp!]
\begin{center}
\includegraphics[width = 5in]{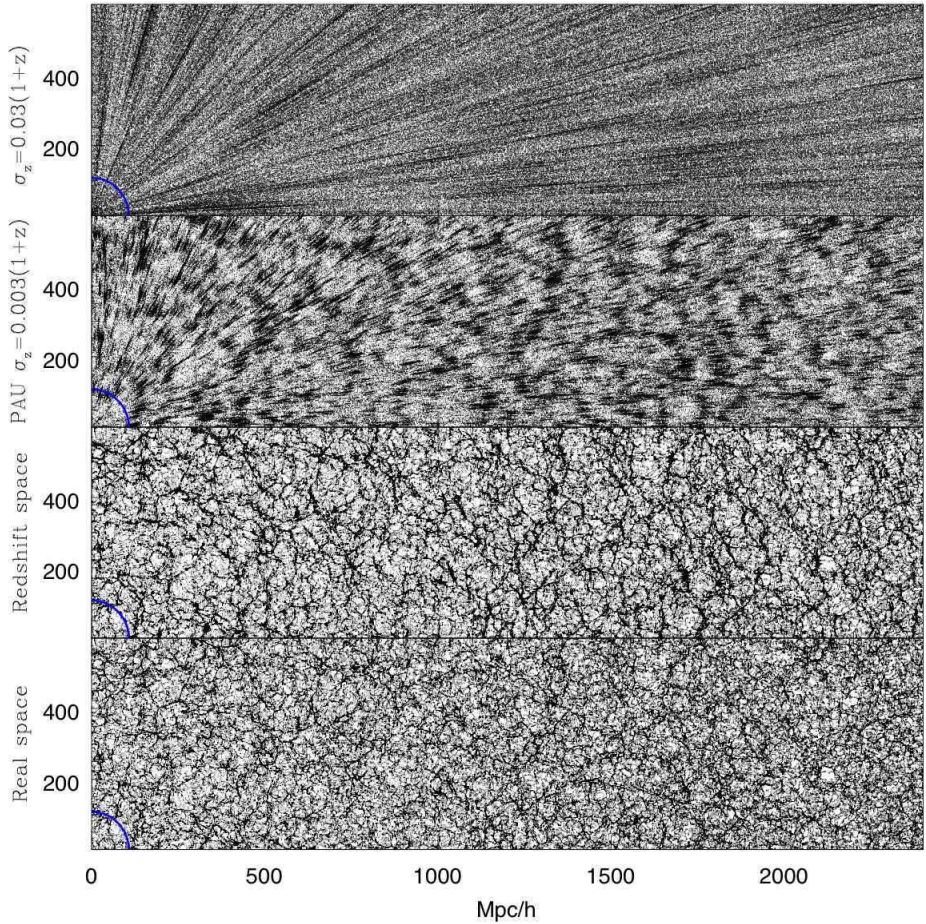}
\caption{The effect of redshift errors on the BAO. A 1$h^{-1}$Mpc thick slice of a simulated dark matter distribution is shown in real space ({\bf bottom panel}) and in redshift space ({\bf second from bottom}). The top two panels are the same distribution in redshift space, but with photometric redshift errors added to each galaxy. The second panel illustrates the effect of Gaussian errors of $0.3\%$ (as expected for the Physics of the Accelerating Universe (PAU) survey), while the top panel has a $3\%$ error. The redshift errors make it difficult to reconstruct the Baryon oscillation scale, which is shown in the bottom left hand corner of all panels (as a circle of radius $100h^{-1}\mathrm{Mpc}).$ From Ben\'{i}tez {\em et al.,} 2008 \cite{pau}.\label{pau_plot}}
\end{center}
\end{figure}

\begin{figure}[htbp!]
\begin{center}
$\begin{array}{@{\hspace{-0.45in}}c@{\hspace{-0.22in}}c}
\includegraphics[width = 3.0in, height = 2.8 in]{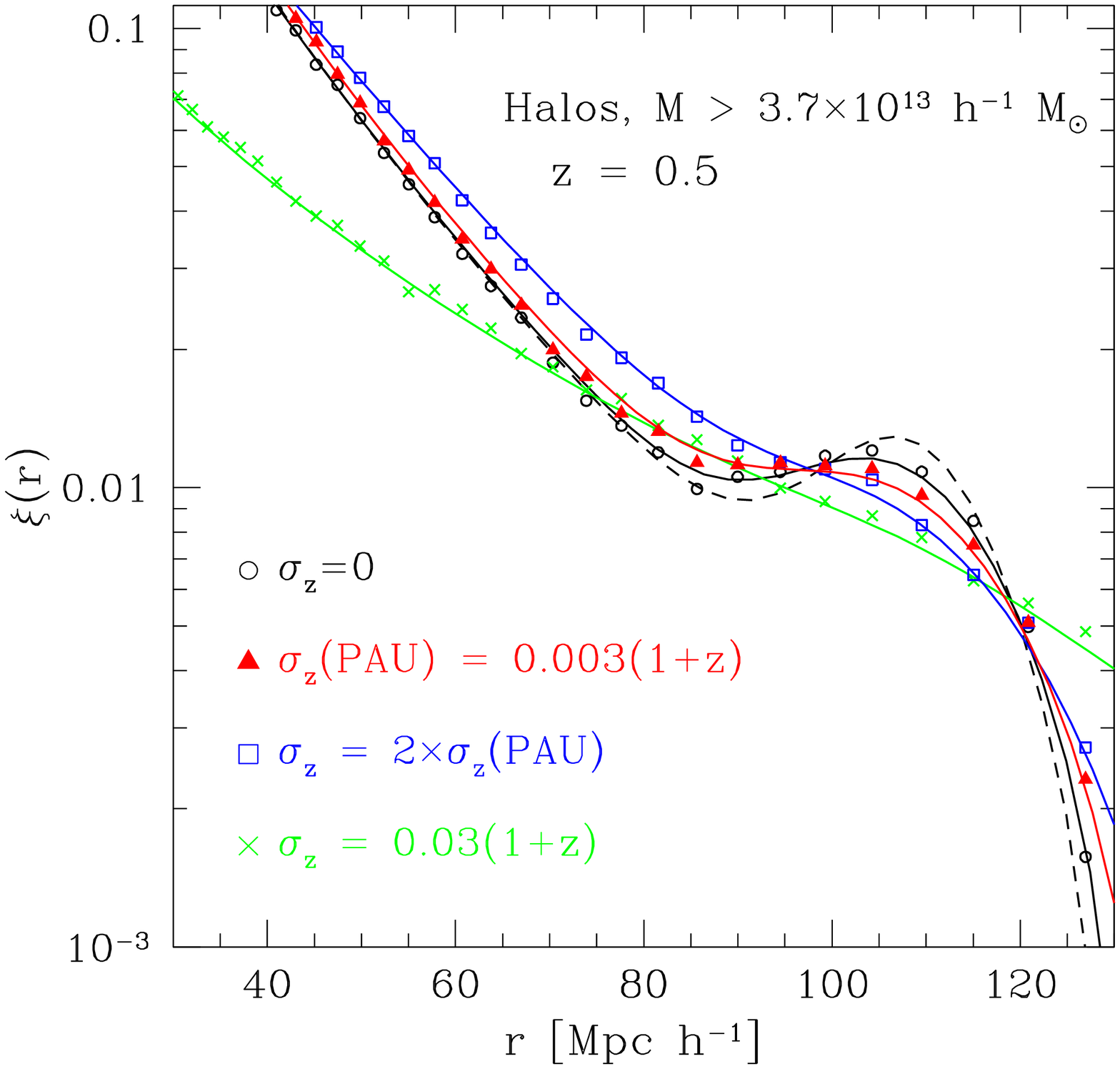} &
\includegraphics[width = 3.0in, height = 2.8 in]{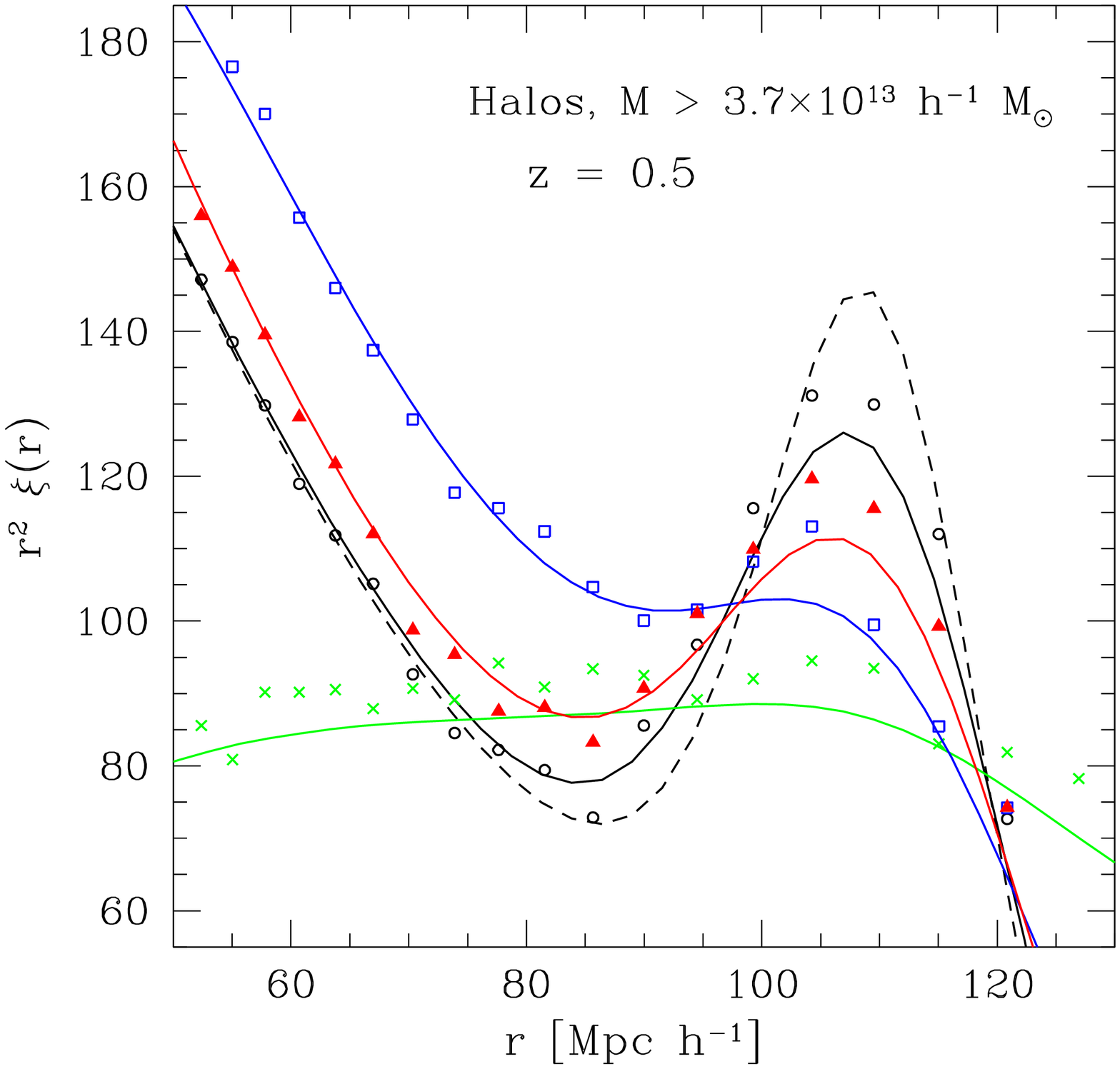}\\ [0.0cm]
 \end{array}$
\caption{The BAO signature smeared by photo-z errors -- the reconstructed correlation function `bump' at $z=0.5$. The circles are the correlation function from a million simulated LRG halos with mass $M>3.7\times10^{13}\, h^{-1}\, M_{\odot},$ from a simulated survey volume of $V=27\, h^{-3}\,{\rm Gpc}^{3}.$ The reconstructed linear correlation function with $b=3$ is shown by the {\bf dashed} line, while the {\bf solid} line shows the nonlinear prediction from renormalised perturbation theory \cite{rpt1}. The {\bf triangle}, {\bf square} and {\bf cross}  symbols show the measured correlation function after a Gaussian error of $\sigma_{z}/(1+z)=0.003,0.007$ and $0.03$ in the line-of-site direction is introduced. The corresponding solid lines are the analytical predictions for the damping of the Fourier space power spectrum from photometric errors using nonlinear corrections. Figure taken from Ben\'{i}tez {\em et al.,} 2008 \cite{pau}. \label{pau:photoz}}
\end{center}
\end{figure}

We can model the radial degradation as a suppression of
the radial power spectrum by a factor $\exp[-(k_{||} \sigma_r)^2]$,
where $\sigma_r$ converts the redshift uncertainty into a physical
distance as per our earlier discussion $\sigma_r = \sigma_z (1+z)/H(z)$. In this sense the effect of photo-z errors are similar to the nonlinear effects discussed later, which wipe out information
about the higher order oscillations. This smearing is clearly visible in Fig.~(\ref{pau_plot}) which shows the progressive smoothing in the radial direction that occurs as the photometric redshift error increases. This smoothing smears out the baryon acoustic peak in the correlation function, as shown in Figure~(\ref{pau:photoz}). Simpson {\em et al.} suggest that in order to recover the dark energy equation of state $w$ to $1\%,$ the dispersion $\sigma_z$ must be known to within $10^{-3}$ \cite{simpson:2009}.

A simulation-calibrated fitting formula for the accuracy with which photo-z surveys can measure the power spectrum is given by Blake {\em et al.} \cite{blake_bao}, which is included in the \name~code \cite{fisher_release}. The key scaling is given by:
\begin{equation}
\frac{\delta P}{P} \propto \sqrt{\frac{\sigma_r}{V}}
\end{equation}
where $V$ is the survey volume and $\sigma_r$ is, as before, the
radial positional uncertainty due to the photo-z error. We see that sacrificing redshift accuracy has a similar effect to
decreasing the survey volume. In particular, Blake and Bridle (2005) \cite{BlakeBridle} estimate that with $\delta z/(1+z) = 0.03$ a
photometric galaxy survey requires approximately an order of magnitude greater area (and hence volume) to match the BAO accuracy achieved by a spectroscopic survey with the same magnitude limit (in the magnitude range $r = 21-23$). Recent claims suggest that there may be
useful information in the small-scale power spectrum which reduces this to a factor five \cite{cai} which would have important implications for the attractiveness of future photometric BAO surveys.

In the limit of large photo-z errors, it is standard to bin the galaxies up into common redshift bins and project the galaxies in each bin onto a sphere at the central redshift of the bin. As a result one considers
instead the angular power spectrum in each bin. This essentially throws away all of the radial information and is therefore maximally conservative \cite{blake_densities, padmanabhan}.
\section{Nonlinear theory}
Our analysis so far has been predicated on the belief that the BAO are reliable standard rulers, which is based on the fact that the BAO scale is $\sim 105~h^{-1}\mathrm{Mpc}$ which, within the context of FLRW models, is squarely in the linear regime since the quasi-linear regime only extends to about $30 h^{-1}\mathrm{Mpc}$, even at $z=0$ (and is significantly smaller at higher redshift). However, every candidate standard ruler or candle has a limit beyond which it cannot be trusted, for either theoretical or observational reasons. In the case of BAO the factors that contribute to the breakdown of confidence are nonlinear clustering and scale-dependent bias. But as we will discuss below, there are reasons to be optimistic even about these potential problems.

A key advantage of the BAO as a cosmological probe is that nonlinearities such as those induced from nonlinear gravitational clustering induce predictable shifts in the oscillation scale and hence can be modelled both analytically and through numerical simulations. The effect of the nonlinearities can then be calibrated for, something which is not possible for many other standard rulers and candles. Here we briefly outline the effects of nonlinearity and techniques to correct for such nonlinearities.

Different prescriptions exist for the method of using the \bao~in the power spectrum as cosmological tools. The full Fourier space galaxy correlation method uses the entire power spectrum (including the shape) \cite{seo2003}, but is sensitive to nonlinearities such as scale-dependent bias and nonlinear redshift space distortions. The effect of these systematics is reduced if one removes the overall shape of the power spectrum by dividing by some reference cosmology \cite{blake_bao, seo_eis07}, however you also lose any information contained in the overall shape and amplitude and so constraints on cosmological parameters will be weaker.
\subsection{Nonlinear bias  \label{nonlinear_bias}}
Measurements of galaxy clustering from redshift surveys yield the galaxy power spectrum, which is traditionally related to the power spectrum of dark matter $P_{\mathrm{DM}}(k),$ (which we are interested in) through the bias $b(k,z)$, \be P_{\mathrm{gal}}(k,z) = b^2(k,z)P_{\mathrm{DM}}(k,z)\,,\ee which in principle can be both redshift and scale dependent \cite{fry,coles, weinberg_bias, hui_parfrey, smith_sheth,sanchez, cresswell}.

Even a moderate scale dependent bias will shift the peaks of the BAO and cause a systematic error in the standard ruler. In the extreme case one could even imprint oscillations not present in the underlying dark matter distribution. Fortunately there is a way out of this degeneracy. Clustering in redshift space is anisotropic due to redshift distortions. The radial component of the galaxy peculiar velocity contaminates the cosmological redshift in a characteristic, scale-dependent manner, which means that the power spectrum in redshift space is not isotropic,
$P(\mathbf{k}) \neq P(k)$. On large scales galaxies falling into overdensities (such as clusters) are `squashed' along the line sight (the Kaiser effect), while on scales smaller than clusters the velocity dispersion of the galaxies within the cluster leads to the `finger of god' effect - clusters appear elongated along the line of sight \cite{hamilton, kaiser}.

We can expand the anisotropic power spectrum as \begin{equation}
P(\mathbf{k},z) = \sum_{l=0,2,4,...} P_l(k,z){\cal L}_l(\mu) \end{equation} where ${\cal L}_l(\mu)$ are the Legendre polynomials, $\mu = \cos(\theta),k=|\mathbf{k}|$ and the monopole $P_0(k,z)$ is the spherically averaged power spectrum we have been discussing for most of this review. The odd moments vanish by symmetry. Studies have shown that the extra information in the higher order moments $P_l$ allow the recovery of essentially all the standard ruler information, even marginalising over a reasonable redshift and scale-dependent bias (e.g. a four-parameter model), with future experiments \cite{yamamoto_bassett}. To understand why the different multipoles would break the bias degeneracy, remember that the amplitude of the redshift distortions is controlled by  the parameter $\beta = \Omega_m^{\gamma}/b,$ where $\gamma \sim 0.6.$ Imagine an observed monopole galaxy power spectrum. If $b \rightarrow 0$ then amplitude of the dark matter power spectrum must increase to leave the galaxy clustering unchanged. The larger dark matter clustering will lead to larger velocities and hence larger redshift distortions which will be visible in the dipole and quadrupole power spectra. Including information from the full power spectrum allows one to calibrate for such a scale-dependent galaxy bias \cite{zhang, percival_white}.

Before moving on we note that the dark matter power spectrum is the product of the initial power spectrum of the Universe, the growth function $G(z)$ (defined in Eq.~(\ref{growthz})) and the transfer function $T(k)$: \be P_{\mathrm{DM}}(k,z) = G^2(z)T^2(k) P_{I}(k) \label{psdm}\,.\ee It is clear from Eq.~(\ref{psdm}) that at the level of the monopole power spectrum the growth function is completely degenerate with the a general bias. Redshift distortions and non-Gaussian clustering (measured e.g. through the bispectrum) offer the opportunity of determining the growth independent of the bias using the same principle we have discussed above.
\subsection{Movement and broadening of the peak}
The main systematic error due to nonlinearity is the shift in the peak of the correlation function due to mode-mode coupling, as has been studied extensively \cite{crocce, smith_sheth, smith_movement}. There are a couple of effects at play here. First, if the broadband correlation function (i.e. the smooth part without the peak) changes with time, the acoustic peak will shift too, simply due to elementary calculus. Secondly, consider the simple physical model introduced in Section \ref{linphys}, where we thought of the correlation function peak as arising from an acoustic wave that moves outwards before stalling at recombination with a galaxy at the origin from which the spherical shell expanded. This relies on the insight that the two-point correlation function, which is a joint probability, can be rewritten as a conditional probability: what is the probability of finding a galaxy at distance, $s,$ {\em given} that there is a galaxy at the origin.
If we then consider the nonlinear evolution of such a sharp density shell, at rest shortly after recombination, we would expect it to undergo some collapse over the history of the cosmos due to its own self-gravity and due to the gravity provided by the galaxy at the centre thereby shrinking the radius of the shell and hence the standard ruler length, by a small but systematic amount of $1-3\%$.

\begin{figure}[htbp!]
\begin{center}
\includegraphics[width = 3.8in, height = 3.2in]{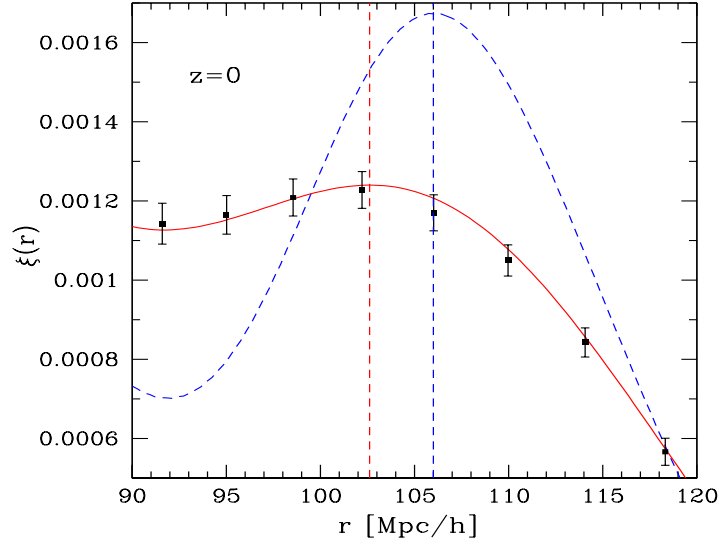}
\caption{Movement of the Baryon Acoustic Peak.  The correlation function at z=0 illustrates how the transfer of power to smaller scales due to nonlinearities, leads to a shift of the BAP in the correlation function, $\xi(r)$. The linear peak, indicated by the {\bf dashed} line is both broadened and shifted towards smaller scales. The {\bf solid} line shows the prediction for the shift from renormalised perturbation theory (RPT) \cite{rpt1, rpt2}. The vertical lines denote the corresponding maxima of the linear and nonlinear correlation functions. From Crocce {\em et al.,} 2008 \cite{crocce}.\label{peak_movement}}
\end{center}
\end{figure}

As can be seen in Fig.~(\ref{peak_movement}), nonlinearities not only shift the peak, but also smooth out and broaden the peak of the correlation function. We can again understand this qualitatively using our simple model. Imagine a galaxy forms on the BAP shell. At the nonlinear level this galaxy is subject to the combined interactions with all other over and under-densities. In any given density realisation this net force may pull the galaxy outwards or inwards. Since the correlation function is computed by averaging over all available galaxies, the average effect is to broaden the BAP. The only constant in all the shells is the galaxy at the centre which causes the small inwards shift of the BAP as described above.

This broadening of the BAP equivalently can be thought of damping the oscillations in the power spectrum\footnote{The correlation function and power spectrum are introduced in Section~\ref{stat}.} on small scales. Broadening the peak obviously makes reconstruction of the position of the peak -  and hence the standard ruler length - less accurate, hence degrading dark energy constraints. Broadening the peak in the correlation function washes out the oscillations in $P(k)$ at large wavenumbers or small scales.

\begin{figure}[htbp!]
\begin{center}
$\begin{array}{@{\hspace{-0.2in}}c@{\hspace{-0.22in}}c}
\epsfxsize=3in
\epsffile{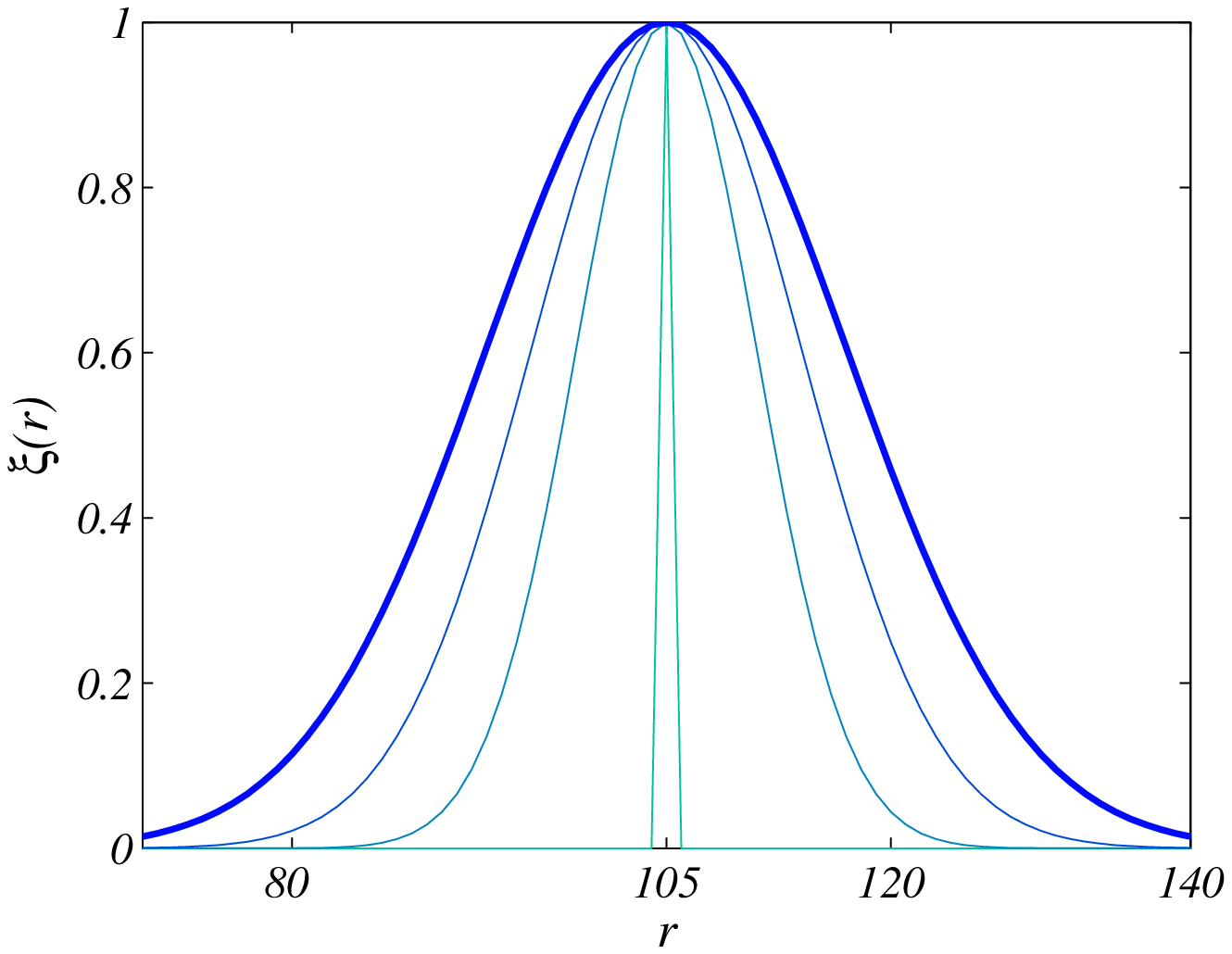} &
	\epsfxsize=3in
	\epsffile{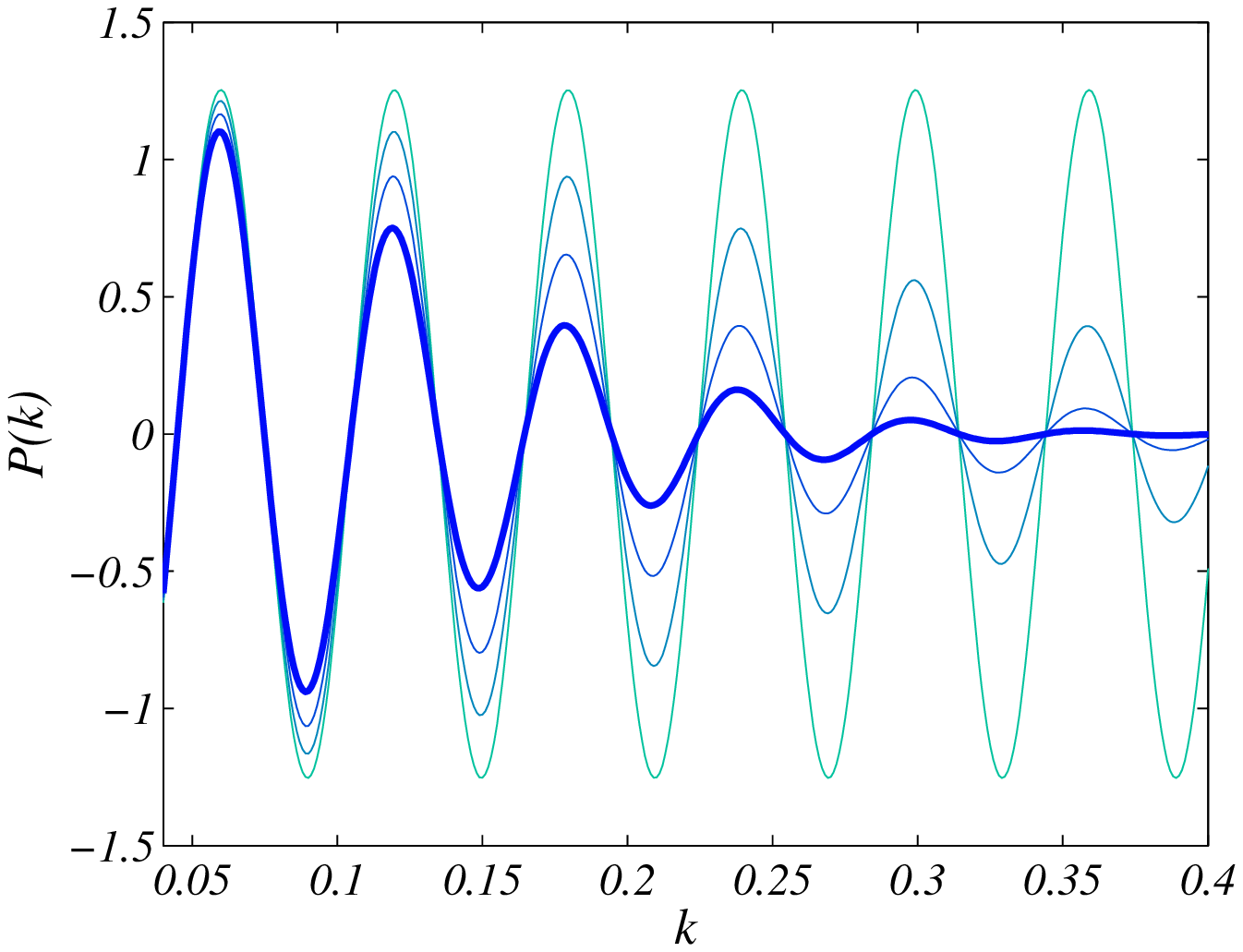}\\ [0.0cm]
 \end{array}$
  \caption{Smoothing out the baryon acoustic signal in the Fourier pair $\xi(r), P(k)$. Increasing the width of the correlation function bump corresponds to the damping of the acoustic oscillations in the power spectrum, particularly severely at large $k$. Both effects make reconstruction of the standard ruler length more noisy. \label{xi_pk}}
  \end{center}
 \end{figure}

\begin{figure*}[htbp!]
\begin{center}
$\begin{array}{@{\hspace{-0.0in}}c}
\includegraphics[width = 3.5in, height = 3.0in]{rings_200_400.eps}\\ [0.0cm]
\includegraphics[width = 3.5in, height = 3.0in]{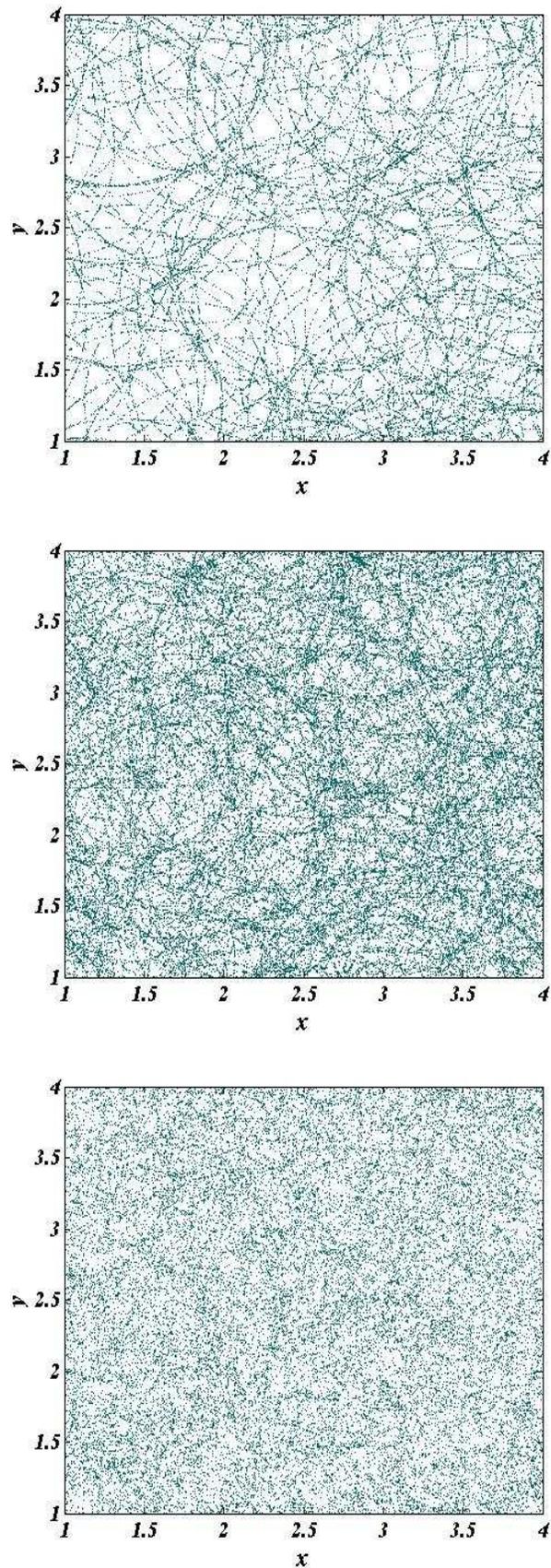}\\ [0.0cm]
\includegraphics[width = 3.5in, height = 3.0in]{rings_875_100.eps} \\ [0.0cm]
 \end{array}$
  \caption{Hiding the characteristic scale. As the peak is broadened from top to bottom as shown schematically in Figure~(\ref{xi_pk}), the underlying rings of power are lost, and must be recovered statistically. c.f. Figure~(\ref{rings:fig}). The number of points are kept the same in each panel. \label{rings:fig_broad} }
  \end{center}
 \end{figure*}

We can illustrate this analytically as follows. Let us model the correlation function as a Gaussian bump shifted so it is centered at a scale $r_*,$ or \be\xi(r) = \exp{\left(-\frac{\left(r-r_*\right)^2}{2\sigma^2}\right)}. \ee
Hence the power spectrum is given by \bea P(k) &=& \int_{-\infty}^{\infty} \exp{(-\frac{\left(r-r_*\right)^2}{2\sigma^2})}\exp{(-ikr)}dr  \nonumber \\
&=&\sqrt{\frac{\pi}{2}}\exp{(-ikr_*)}\exp{\left(-k^2\sigma^2\right)}\eea  Figure~(\ref{xi_pk}) illustrates this toy-model correlation function consisting of a Gaussian shifted to some preferred scale $r_*$, and the corresponding power spectrum $P(k).$ The oscillations are given for a range of widths of the Gaussian bump, $10< \sigma< 35.$ Clearly as the Gaussian broadens, the oscillations in the power spectrum are washed out, making their detection harder. Recalling the illustration of rings of power in Figure~(\ref{rings:fig}), we can examine the effect of successively broadening the rings from which the points are drawn. This this shown in Figure~(\ref{rings:fig_broad}) which shows the smearing of the characteristic radius implying an increased error in the standard ruler measurement.
\subsection{Reconstruction}
While non-linear gravitational collapse broadens and shifts the peak of the correlation function, Eisenstein {\em et al.} \cite{eisenstein_nonlin} point out that the map of galaxies used to extract the power spectrum in redshift space can also be used to map the velocity field. Since the galaxies are essentially test particles in the standard $\Lambda$CDM paradigm, this velocity field can then be used to undo the effects of the nonlinear clustering or equivalently to reconstruct the position and sharpness of the linear acoustic oscillation peak by moving densities to where they would have been had linear theory held at all times. By considering a pair of galaxies separated by the characteristic BAO scale, Eisenstein, Seo and White (2007) \cite{ESW} show that the majority of the corrupting signal comes from wavenumbers $k \sim 0.02-0.2 h \mathrm{Mpc}^{-1}$. Larger wavelengths coherently move both galaxies while smaller scales are weak because the power spectrum has little power there. The typical distances induced by nonlinear corrections are around $10h^{-1}\mathrm{Mpc}$.

\begin{figure}[htbp!]
\begin{center}
	\includegraphics[width = 3.5in, height = 3.0in]{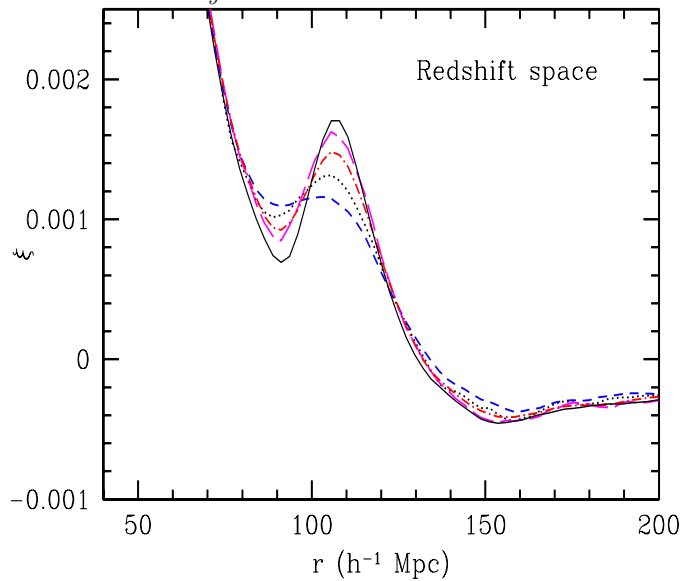}
  \caption{Reconstructing the Baryon Acoustic Peak (BAP) - The nonlinear effects generate bulk flows that can be reconstructed from the galaxy distribution itself hence allowing the nonlinear movements of densities to be undone to good accuracy, thereby both sharpening the acoustic peak and moving it to the correct linear position.  From Eisenstein {\em et al.,} 2007 \cite{eisenstein_nonlin}.
\label{recons}}
  \end{center}
 \end{figure}
 Various methods can be followed to reconstruct the velocity and density field which are summarised in \cite{eisenstein_nonlin}. Eisenstein {\em et al.} move the measured densities back to their linear locations using the following prescription. First they smooth the density field on about $10h^{-1} \mathrm{Mpc}$ scales. Then they compute the Lagrangian displacement field, $\overrightarrow{q}$ which is assumed to be irrotational (no vector perturbations) and which obeys the linear prediction $\Delta\cdot\overrightarrow{q} = -\delta$. All particles are then shifted by $-\overrightarrow{q}$. In redshift space the densities are then boosted by $1 + d(\ln G)/d(\ln a),$ where $G$ is the growth factor, to account for the linear redshift distortions.

Figure~(\ref{recons}) shows the reconstruction of the acoustic peak in the real-space correlation function using these techniques with improvement in the accuracy of the peak by a factor of 2 to 3, with similar results for the redshift-space correlation function.

 The impact of reconstruction on BAO survey optimisation is still to be explored but an immediate implication is that low-redshift surveys, where the nonlinearities and broadening of the peak is stronger, will benefit more from reconstruction than high-redshift surveys where the effect of nonlinearities have not had time to imprint on the relevant scales. This will, in turn, make low-redshift BAO surveys more interesting, allowing smaller telescopes (e.g. 2-4m class) to compete with the larger telescopes (8-10m class) in the BAO stakes. We now move on to a discussion of targets for BAO surveys.
\section{Target Selection}

\begin{figure}[htbp!]
\begin{center}
\includegraphics[width=2.5in,height = 3.5in, angle = 270]{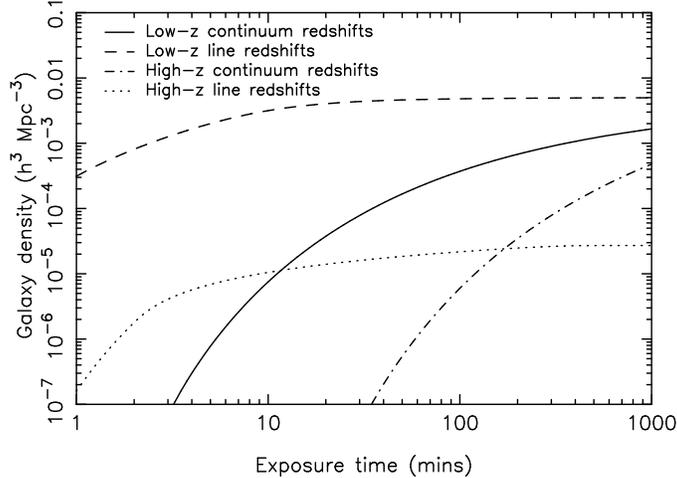}
\caption{Exposure time required to reach a given galaxy density on an 8m class telescope for four typical BAO targets: red galaxies at $z\sim1$ {\bf (solid line)} where the continuum is used to obtain  redshifts; blue galaxies at $z\sim1$ {\bf (dashed line)} where redshifts are obtained using galaxy emission lines; red galaxies at high redshift ($z\sim3$ -- {\bf dot-dashed line}) and blue galaxies again at high redshift {\bf(dotted line)}. The plateau in the galaxy density occurs when the surface density reaches the spectroscopic fibre density (a single pointing of the telescope is assumed). Figure from Parkinson {\em et al.,} 2006 \cite{parkinson06} for the WFMOS survey. \label{integration}}
\end{center}
\end{figure}

A key decision in undertaking any BAO survey is the choice of target, since the bias, $b$, of different potential targets differ considerably as a function of morphology, colour etc... relative to the underlying dark matter distribution, in addition to any redshift or scale dependence; $P_{target}(k) = b^2 P_{DM}(k)$. This translates into different optimal target densities since if one requires $n P_{target} \sim 1$ (the criterion translates to $n \sim 1/(b^2 P_{DM}))$: one needs fewer more highly biased tracers of clustering than weakly biased targets.  For example, Luminous Red Galaxies (LRGs) are highly biased tracers, $b_{LRG} \sim 1.5-2$ \cite{tegmark_lrg} since they are typically found in clusters while blue spirals are typically field galaxies and hence are not strongly biased. All modern studies of BAO use dedicated targets, the choice of which typically trades off bias versus integration time.  Integration times for various types of possible targets for a large 5000 fibre 10m class survey (such as WFMOS) are shown in Fig.~(\ref{integration}).

\subsection{Luminous Red Galaxies (LRGs)}
\begin{figure}[htbp!]
\begin{center}
$\begin{array}{@{\hspace{-0.2in}}c@{\hspace{-0.22in}\vspace{-0.22in}}c}
\includegraphics[width=3.0in, height = 3.0in]{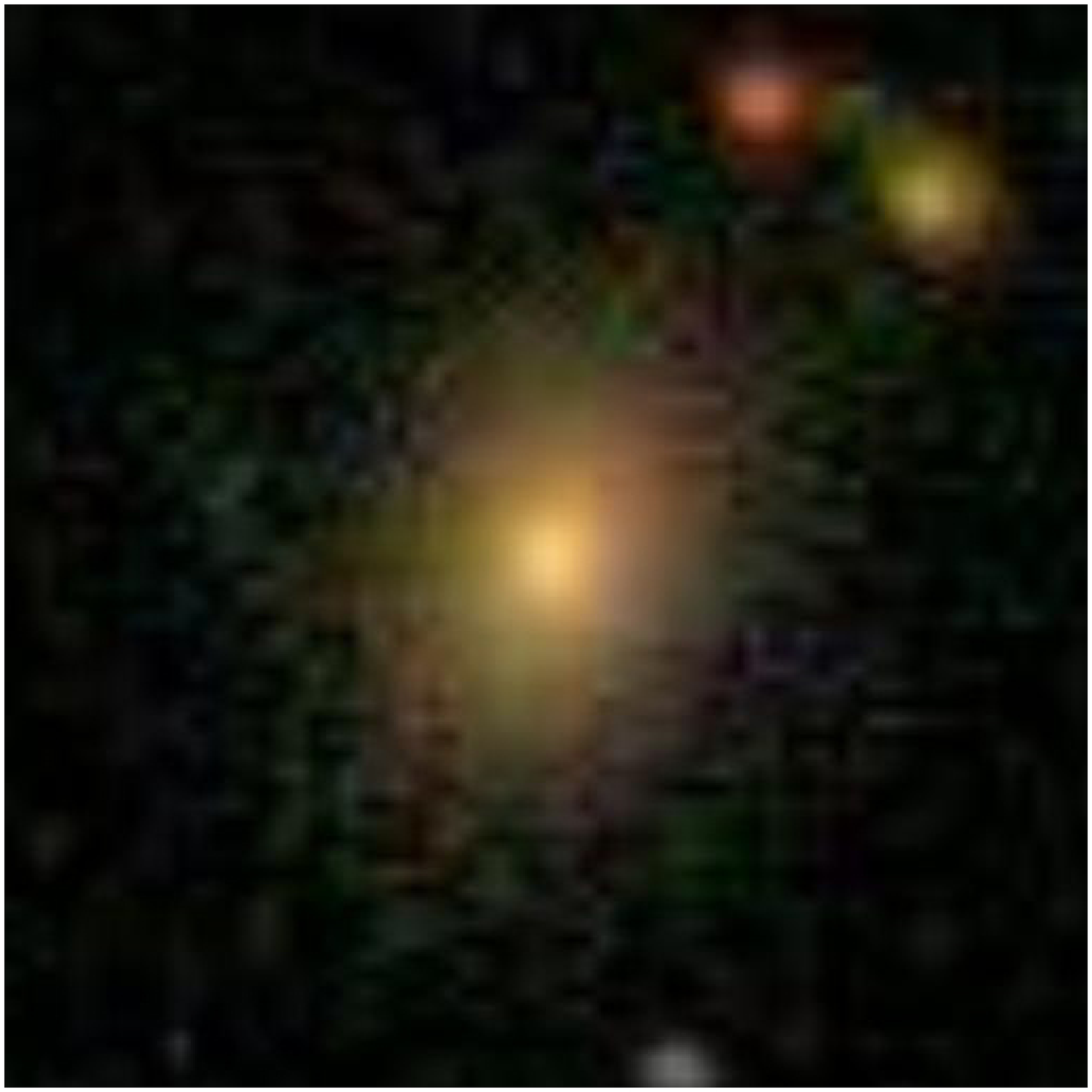} &
\includegraphics[width=3.0in, height = 2.8in]{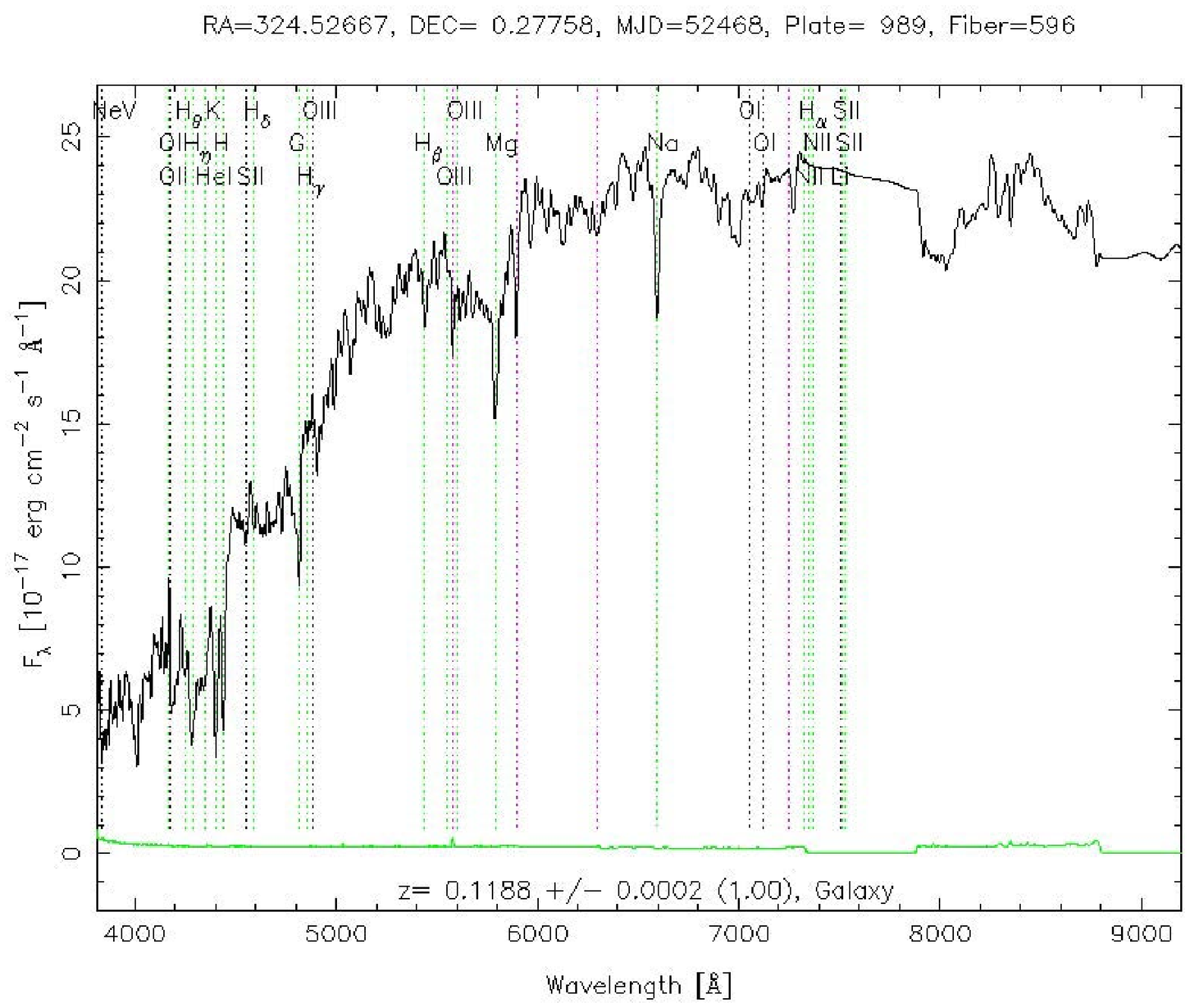}\\ [0.0cm]
\end{array}$
  \caption{Example of a Luminous Red Galaxy. The Sloan Digital Sky Survey DR6 image {\bf (top panel)} and spectrum {\bf (bottom panel)} of a typical LRG and showing a lack of emission lines, which means that the continuum must be measured to obtain a redshift. From http://skyserver.sdss.org/ \label{sdss_lrg} }
  \end{center}
 \end{figure}
Luminous Red Galaxies rose to prominence with the SDSS LRG survey \cite{eisenstein_05}. They are typically ``red and dead," passive elliptical galaxies with featureless spectra. A high S/N LRG spectrum is shown in Figure~(\ref{sdss_lrg}). The redshift is derived from the position of the $4000$ \A break which therefore requires long integration times even on a 10m class telescope, for redshifts $z >1$. This is counteracted by the large bias of LRGs which means that the required target density is significantly lower. The latter advantage, plus the ability to efficiently find LRGs in optical photometric surveys like the SDSS survey has lead to LRGs been chosen as the targets for the BOSS SDSS-II Survey\footnote{www.sdss3.org} \footnote{http://sdss3.org/collaboration/description.pdf}.
\subsection{Blue galaxies}
\cite{wigglez}.
\begin{figure}[htbp!]
\begin{center}
$\begin{array}{@{\hspace{-0.22in}}c}
\epsfxsize=4.5in
\epsffile{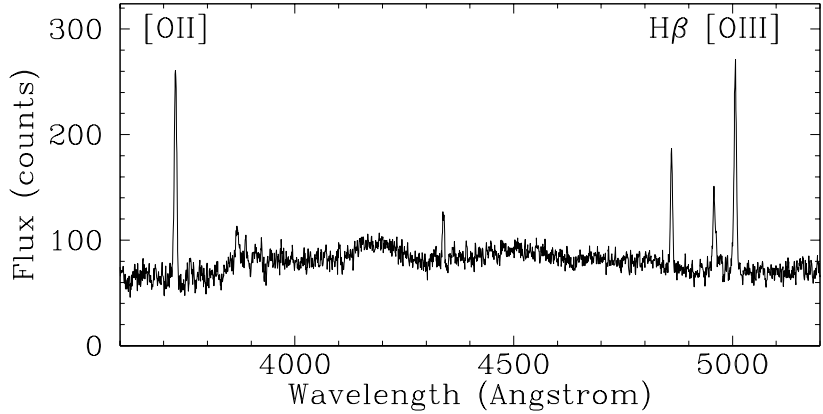} \\ [0.0cm]
 \end{array}$  \caption{A star-forming galaxy spectrum showing the characteristic emission lines used for redshift determination taken as part of the
WiggleZ survey \cite{wigglez}. \label{fig:blue_gal}}
  \end{center}
 \end{figure}
While redshifts for LRGs are obtained from the continuum spectrum, blue, star-forming galaxies have strong emission lines which provide good redshifts (see Fig.~(\ref{fig:blue_gal})) from for example the OII doublet at 3727 $\A$ which is within the optical band at redshifts $z < 1.4$ -- expected integration times for this line using a 10m class telescope are around 15 minutes. Despite this, the low bias means that a much higher target density is required compared to LRGs. Selection of star-forming targets is achieved with a combination of optical and UV imaging and forms the basis for the WiggleZ survey which uses a combination of SDSS and GALEX (UV) imaging for selection.
 \subsection{Lyman Break Galaxies}
The standard emission lines go out of the optical band at $z \sim 1.4$ leading to the redshift desert for optical surveys because of a dearth of emission lines at wavelengths $< 3000 \A$. This ``drought" is broken by Ly-$\alpha$ at the wavelength of $1216 \A$ which comes into the optical passbands around a redshift of $z \sim 2.3$ and remains there until $z \simeq 6.4,$ making it an ideal target at high redshift. For galaxies at higher redshifts the Ly$-\alpha$ break moves into different bands and the galaxy will have negligible flux in (for example) the U band, but strong flux in the V band - hence the UV `drop-out', multi-colour imaging of the galaxy is hence used to photometrically determine the redshift of the galaxy. Lyman Break Galaxies (LBG) take long integration times (see the high-$z$ continuum curve in Fig.~(\ref{integration})) but there are large numbers of them.
\subsection{Lyman Emitting Galaxies}
A small set of LBGs also have strong Ly-$\alpha$ emission lines. When they exist they provide ideal targets for redshifts due to the strong emission, however their number density is somewhat unknown. They are the target of preference for the HETDEX\footnote{http://hetdex.org} \cite{hetdex1} instrument planned for the Hobby-Eberly Telescope in the redshift interval $1.9 < z < 3.5$, which is expected to detect $\sim8\times10^5$ of these Ly-$\alpha$ emitting galaxies.
\subsection{Supernovae}
LSST will potentially detect millions of photometric Type Ia supernovae (SNIa). Zhan {\em et al.} have proposed that these supernovae could be used to measure BAO \cite{zhan} at $z < 1$. Hence the same data could provide both $d_L$ and $d_A,$ providing constraints that are insensitive to Cosmic Microwave Background priors. The advantage of SNIa as  photometric BAO targets over galaxies is that the photometric redshift error is typically significantly smaller ($\sigma_z \sim 0.02$) due to the well-sampled, multi-epoch spectral templates that will be available from current and future low-z supernova surveys.
\subsection{Lyman Alpha Forest}
Sampling the underlying dark matter distribution at a discrete set of $N$ points makes it difficult to uncover subtle underlying patterns due to shot noise. Instead, a potentially superior method would be to take 1-dimensional slices through the density distribution. This is the idea behind using the Lyman-alpha forest to probe the BAO. McDonald and Eisenstein (2007) \cite{lyman_1} discuss such a survey at redshift $2.2 < z < 3.3,$ and project constraints on the radial and tangential oscillation scales of order $1.4\%$. Such a survey could be performed at the same time as a spectroscopic galaxy redshift survey, which is the plan for the BOSS SDSS-III survey.
\subsection{21cm Neutral Hydrogen}
One step better than a 1-D slice is a full 3-D slice through the neutral hydrogen distribution. This is the possibility afforded by neutral hydrogen surveys based on the 21cm HI emission line \cite{Blake,abdallaBAO, wyithe_loeb_geil,chang_21}. The advantage of this probe is that neutral hydrogen should be ubiquitous at all redshifts, although the precise redshift dependence of the HI density is unknown and is further affected by uncertainties in cosmic reionisation. However, in principle, HI surveys will be able to probe deep into the dark ages before the formation of galaxies, providing access to the cosmic density field uncontaminated by nonlinearities.
\section{Current and Future BAO Surveys\label{experiments}}
The key elements for a BAO survey are redshift accuracy, redshift coverage, area and volume (of course the latter three are not independent at fixed total survey time). The ideal instrument therefore has large field of view (the area it can see at any one time), large mirror size allowing short integration times and if it is taking spectra, the ability to harvest large numbers of spectra simultaneously. For future surveys being considered now, this means fields of view in excess of $1\deg2$, mirrors greater than 4m in size and for spectroscopic surveys, the ability to take at least $1000$ spectra simultaneously, using a multi-fibre or other technology.
\subsection{Spectroscopic Surveys}
We now discuss in rough chronological order the current and future spectroscopic BAO surveys\footnote{These would be called Stage II, III and IV surveys in the DETF report \cite{detf}.}. First up are the final Sloan Digital Sky Survey (SDSS-II) LRG and main galaxy surveys at $z \simeq 0.35$ and $z \simeq 0.1$ respectively and which will cover about $10,000\deg2$ in the northern hemisphere. Next is the WiggleZ survey using the 400 fibres on the AAT \cite{wigglez} and covering $1000\deg2$ over the redshift range $0.2<z<1.0$, with a median redshift of $z = 0.6$, which will be completed in 2010 and will measure $H(z)$ and $d_A(z)$ to around $5\%$. Beyond that is the Baryon Oscillation Spectroscopic Survey (BOSS), part of the SDSS-III \footnote{www.sdss3.org} survey and scheduled to operate over the period 2009-2014. The relatively small diameter (2.5m) of the SDSS telescope combined with the large field of view means that BOSS will focus on a wide-and-shallow survey measuring approximately 1.5 million LRGs at $z \leq 0.7$ and around 160,000 Ly-$\alpha$ forest lines at $2.2 < z < 3$ and giving projected absolute distance measurements of $1.0\%$ at $z = 0.35$, $1.1\%$ at $z = 0.6$, and $1.5\%$ at $z = 2.5$ \footnote{http://sdss3.org/collaboration/description.pdf}.

Projects on a similar, 2010-2015, time-scale include FMOS and LAMOST. FMOS is an infra-red spectrograph for Subaru with $400$ fibers which could undertake a moderate but interesting BAO survey in the redshift desert at $z \sim 1-1.7$ over $\sim 300\deg2$\footnote{http://www.sstd.rl.ac.uk/fmos/} while LAMOST is the Chinese 4m telescope with a $4000-$fiber spectroscopic and $20\deg2$ field of view which should enable a very effective BAO survey similar to BOSS both at $z \sim 1$ and high-$z$ using quasars\footnote{http://www.lamost.org/en/} \cite{lamost_paper}.

Another exciting BAO survey is the Hobby-Eberly Dark Energy eXperiment (HETDEX)\footnote{http://hetdex.org} \cite{hetdex1} which will target the highly biased Ly-$\alpha$ emitting galaxies over the range $1.8 < z < 3.7.$ Such a survey over $200\deg2$ would probe about $5 \mathrm{h}^{-3}\mathrm{Gpc}^3$ with approximately one million galaxies, allowing HETDEX to provide $\sim 1\%$ measurements of $d_A(z)$ and $H(z)$ at three redshifts over the survey range. An attractive feature of HETDEX is that it does not need any pre-selection imaging; targets are acquired purely by chance using integral field spectrographs \cite{hetdexde}.

The next major advance in the spectroscopic BAO domain would be enabled by the Wide Field Multi-Object Spectrograph (WFMOS) on a 10m class telescope, such as Subaru \cite{WFMOS_feasibilitystudy,wfmos1}. While WFMOS has been cancelled as a Gemini project, it may still take place in a different form. The default plan for WFMOS called for a large field of view ($> 1\deg2$) and a large number of fibers (at least $3000$, although the optimal number is still being investigated \cite{parkinson06}). Slated for a $\sim$ 2015 start, the default WFMOS-like surveys would measure in excess of one million LRGs or blue galaxies at $z = 0.8-1.3$ over an area of $2000-6000\deg2$ and of order one million Lyman Break Galaxies at $z = 2.5-4$ over a somewhat smaller area, providing percent level measurements of both $d_A(z)$ and $H(z)$ at $z=1$ and $z=3$ \cite{eisenstein2003,seo2003, parkinson06, parkinson09}.  The high-$z$ component of the survey would provide a key leverage against uncertainties in curvature and $w(z)$ at $z > 1$. It would also be a powerful probe of modified gravity \cite{yamamoto_bassett} and allow high spectral resolution archeology of the Milky Way to understand the origins of its stellar populations \cite{WFMOS_feasibilitystudy}.

Beyond the 2015 timescale there are a number of planned and proposed missions in various stages of preparation. Perhaps the simplest proposal is the BigBOSS ground-based experiment which, over a ten-year period would be able to compete with Dark Energy Task Force Stage IV experiments \cite{bigboss}. Building on BOSS it would use 4m telescopes at both northern and southern NOAO sites (initially 6 years at KPNO potentially followed by four years at CTIO after the Dark Energy Survey) fitted with a new 4000-fiber R=5000 spectrograph covering $340-1130$nm with a $7\deg2$ field of view, yielding a survey sample of up to 50 million galaxies and a million quasars over $24000\deg2$ and allowing superb measurements of the BAO and redshift space distortions in the range $0.2< z < 3.5$. At low-z, targets would be LRGs while for $1 < z < 2$ BigBOSS would target bright OII emission line galaxies with the QSOs taking over at $z > 2$.

A more radical proposal is to use slitless spectroscopy \cite{BOP, space_slitless} which is one possibility for the spectroscopy component of the proposed EUCLID survey, which is a combination of the earlier SPACE \cite{space} and DUNE \cite{dune} missions. The spectroscopic component posits an all-sky near-IR survey down to H=22 which would provide of order 150 million redshifts. EUCLID would aim for launch around 2018 if it is chosen as the winner of the ESA Cosmic Visions program.  A further space BAO proposal for the DOE-NASA JDEM mission is ADEPT which would also gather around 100 million redshifts over the redshift range $z < 2$. Recently, the possibility of a JDEM-EUCLID merger has been raised due to the obvious complementarity of the science of the two programs and budgetary constraints, although the technical and organisational challenges of building such a complex joint US-Europe mission are likely to be significant.

A very different direction is provided by radio BAO surveys. Despite the inherent weakness of the 21cm signal it is likely that radio telescopes will play an important role in future cosmology. This is primarily driven by the fact that the sensitivity of radio telescopes for projects such as BAO scales as the square of their area, unlike optical telescopes whose sensitivity scale linearly with their diameter. This, together with technologies such as synthetic aperture arrays that allow very large fields of view offer the appealing possibility of surveying huge volumes at very high target densities.

An exciting proposal in this direction is the 21cm Hubble Sphere Hydrogen Survey (HSHS) \footnote{http://h1survey.phys.cmu.edu/} which would measure the BAO in neutral hydrogen over the whole sky out to $z = 1.5$. This highly ambitious proposal would provide essentially cosmic variance-limited measurements of the power spectrum in bins of width $\Delta z \sim 0.1$ and exquisite accuracy on $d_A$ and $H(z)$ in the same bins. The key to the HSHS concept is simultaneously combining huge collecting area with very a large field of view. This can be achieved, at what is hoped to be low cost, by using multiple fixed parabolic cylinders which provide drift scans of the entire sky everyday. In this sense, one of the Fourier transforms needed to form an image is undertaken in software (`along the cylinder') while the other is done in hardware (`in the parabolic direction'). HSHS is unusual for a galaxy survey because of its low angular resolution of around $1',$ adapted for statistical analysis of the BAO rather than producing a galaxy catalogue as its primary output. In this sense HSHS resembles a CMB experiment for neutral hydrogen.

A more ambitious proposal is that of the full Square Kilometer Array (SKA) \footnote{http://www.skatelescope.org/} which may be a fully software telescope at 21cm frequencies, with both Fourier transforms being done in software and using completely flat reflectors. The great advantage of such purely synthetic apertures would be that detectors would essentially see all of the visible sky all of the time, providing the ultimate field of view \cite{ska_flat}. This idea appears to have been rediscovered in the form of the Fast Fourier Transform Telescope \cite{FFTT}. The SKA would provide essentially cosmic variance limited BAO measurements out to $z=1.4$ and beyond with of order $10^9$ redshifts, but with sub-arcsecond angular resolution, allowing in addition excellent weak lensing measurements \cite{Blake}. While SKA will be an exceptional BAO machine, pathfinders leading up to the full SKA will also provide the first detections of the BAO in the radio \cite{blake09}.

Beyond SKA one can imagine using radio surveys to probe the BAO at very high redshifts, $z > 10$, where many more modes are in the linear regime. Since there are essentially no galaxies above this redshift, neutral hydrogen will likely be the only way to test dark energy in the dark ages at \cite{cooray_ska}.
\subsection{Photometric Surveys}
Spectra are slow and expensive to obtain and it is tempting to try to study the BAO
with only multi-band imaging. A large number of photometrically harvested galaxies might
provide a useful probe of
the Baryon Acoustic Oscillations, provided the photometric redshift error is
small enough, as discussed in Section~\ref{sec:photz}.

The current state of the art of photometric redshift surveys is
provided by the MegaZ and related catalogues \cite{megaZ, blake_densities, padmanabhan} based on the SDSS photometry. These catalogues
target LRGs and typically achieve $\delta z \simeq 0.03(1+z)$ redshift
accuracy with approximately $1\%$ contamination from M-star
interlopers after suitable cuts. Although they include more than 1 million LRGs out to $z
\sim 0.7$ and cover 10,000$\deg2$ they do not detect the BAO
with any significance due to projection effects arising from the photometric redshift errors.

Beyond SDSS there are a number of exciting photometric surveys. SkyMapper will essentially provide Sloan in the southern hemisphere
\cite{skymapper} while the Dark Energy Survey (DES) \footnote{https://www.darkenergysurvey.org/} \cite{DES} will use $30\%$ of the 4m CTIO telescope time to cover around $5000\deg2$ and detect of order 300 million galaxies in the five
Sloan photometric bands, $u,g,r,i,z$ over the redshift range $0.2 < z
< 1.3$ while the PS1 phase of the Pan-STARRS project could cover
$3\pi$ steradians of the sky and detect of order 100 million LRGs,
again in five, slightly redder, passbands \cite{cai}. Both surveys
should provide compelling BAO detections in addition to the wealth of
other science including lensing and a rich SNIa dataset.

A further interesting hybrid is the Physics of the
Accelerating Universe (PAU) proposal which one might call an
ultra-photometric or quasi-spectroscopic survey. PAU plans to bridge the gap
between standard photometric and spectroscopic surveys through the use of an order of magnitude more filters than SDSS, DES or Pan-STARRS. Using 40 narrow-band and two broad-band filters covering the
optical range, the aim is to identify the $4000 \A$ break with enough
spectral resolution to determine redshifts to an accuracy of $\delta z
\simeq 0.003 (1+z)$ which is hoped will provide sufficient accuracy to reconstruct the BAO
scale in the radial direction and hence obtain $H(z)$ information as
well as $d_A(z)$ \cite{pau}. Again LRGs are the targets of choice due to their simple spectra and with a survey area of order $8000\deg2$ the desire is to measure such ultra-photometric
redshifts for over $10^7$ LRGs at $z < 1$, although doubts have been raised as to whether this approach is competitive with spectroscopic BAO surveys \cite{pau2}. The correlation function from simulated galaxy halos is given in Figure~(\ref{pau:photoz}), from \cite{pau}. The smearing of the acoustic feature is clearly visible as the photometric redshift error increases.

Beyond these surveys the Large Synoptic Survey Telescope (LSST)
will likely provide the definitive photometric survey for the next two
decades. Covering $20,000\deg2$ of the sky visible from Chile,
LSST would detect every galaxy visible in the optical down to a
co-added limiting magnitude of $r=27.5$, or about 10 billion galaxies. With science operations slated to begin in 2015 or soon thereafter, LSST will yield exquisite detections of the angular BAO as a function of
redshift, albeit without the radial information provided by
spectroscopic or ultra-photometric surveys \cite{LSST}.
\section{Conclusions}
In the era of precision cosmology, standard rulers of ever-increasing accuracy will provide powerful constraints on dark energy and other cosmic parameters. The Baryon Acoustic Oscillations are rooted primarily in linear physics with nonlinearities that can be well-modelled and corrected for. As a result the characteristic scale of these `frozen relics' imprinted into the cosmic plasma before decoupling will likely remain as the most reliable of the Statistical Standard Rulers in the coming decade.
\section{Acknowledgements}
 We would like to thank Chris Blake for detailed comments and Daniel Eisenstein, Yabebal Fantaye, Jacques Kotze, Roy Maartens, Will Percival, Varun Sahni and Alexei Starobinsky for insightful discussions. We thank Txitxo Ben\'{i}tez, Martin Crocce, Daniel Eisenstein, David Parkinson, Will Percival, Kevin Pimbblet, Roman Scoccimarro and Max Tegmark for permission to reproduce figures in this review. BB thanks his WFMOS Team A colleagues and in particular Chris Blake, Martin Kunz, Bob Nichol and David Parkinson for their collaborations and discussions over the years, and the ICG, Portsmouth and the Perimeter Institute for hospitality during his visit during which part of this work was completed. RH would like to thank Princeton University for hospitality and acknowledges support from the NSF PIRE grant OISE/0530095BB during her visit there. Finally we thank Pilar Ruiz-Lapuente for organising the Key Approaches to Dark Energy conference in Barcelona which lead to this review. We acknowledge funding from the NRF, Royal Society and SA SKA while RH acknowledges funding from the Rhodes Trust.

\end{document}